\documentclass[showpacs,amsmath,amssymb]{report}

\usepackage{hyperref}

\usepackage[dvips]{graphicx} 

\begin{document}

\title{Inter-universal entanglement\footnote{Chapter written for the book ``Open Questions in Cosmology", Edited by: Gonzalo J. Olmo, InTech Publishing, Rijeka, Croatia, (2012), ISBN 978-953-51-0880-1. The properly formatted version can be freely downloaded from the \href{http://www.intechopen.com/}{publisher's website}.} }
\author{Salvador J. Robles-P\'{e}rez\\ {\small Colina de los Chopos, Centro de F\'{\i}sica ``Miguel Catal\'{a}n'', Instituto de F\'{\i}sica Fundamental,}\\ {\small Consejo Superior de Investigaciones Cient\'{\i}ficas, Serrano 121, 28006 Madrid (SPAIN) and} \\ {\small Estaci\'{o}n Ecol\'{o}gica de Biocosmolog\'{\i}a, Pedro de Alvarado, 14, 06411-Medell\'{\i}n, (SPAIN).}}

\date{\today}

\maketitle

\begin{abstract}
Quantum information theory and the multiverse are two of the greatest outcomes of the XX century physics. The consideration of entanglement between the quantum states of two or more universes in a multiverse scenario provides us with a completely new paradigm that opens the door to novel approaches for traditionally unsolved problems in cosmology. More precisely, the problems of the cosmological constant, the arrow of time and the choice of boundary conditions, among others. It also encourages us to adopt new points of view about major philosophical ideas. In this chapter, we shall present the main features that may characterize inter-universal entanglement and it will be addressed the customary problems of cosmology from the new perspective that the quantum multiverse scenario supplies us with.

In summary, the appropriate boundary condition that has to be imposed on the quantum state of the whole multiverse allows us to interpret it as made up of entangled pairs of universes. Then, a quantum thermodynamical description of single universes can be given and it can be shown that it may induce observable effects in the energy properties of the Universe. The effects that the boundary condition of the multiverse has on the vacuum energy and the arrow of time of single universes are also studied. As a consequence of inter-universal entanglement, the former might be discriminated from observational data and the latter would favor the growth of cosmic structures that increase the amount of local entropy mainly in the very early phase of the universe. All these characteristics of inter-universal entanglement would eventually impel us to develop the concept of the physical multiverse, one  for which the theory could be not only fallible but also indirectly observed. 
\end{abstract}


\addtocounter {chapter} {1}

\tableofcontents
\section{Introduction}

The concept of the multiverse changes many of the preconceptions made in the physics and cosmology of the last century, providing us with a new paradigm that has inevitably influence on major philosophical ideas. The creation of the universe stops being a singled out event to become part of a more general and mediocre process, what can be thought of as a new ``Copernican turn'' in the natural philosophy of the XXI century. The multiverse also opens the door to new approaches for traditional questions in quantum cosmology. The origin of the universe, the problem of the cosmological constant and the arrow of time, which would eventually depend on the boundary conditions that are imposed on the state of the whole multiverse, challenge us to adopt new and open-minded attitudes for facing up these problems.

It would mean a crucial step for the multiverse proposals if a particular theory could make observable and distinguishable predictions about the current properties of our universe. That would bring the multiverse into the category of a physical theory at the same footing as any other. Then, once the concept of the multiverse has reached a wider acceptance in theoretical cosmology, it is now imperiously needed to develop a precise characterization of the concept of a physical multiverse: one for which the theory could be not only falseable but also indirectly tested, at least in principle. Some claims have been made to that respect \cite{Mersini2008, PFGD2011, PFGD2012}, although we are far from being able to state the observability of any kind of multiverse nowadays.

In order to see the effects of other universes in the properties of our own universe, it seems to be essential  considering any kind of interaction or correlation among the universes of the multiverse. Classical correlations  in the state of the multiverse would be induced by the existence of wormholes that would crop up and connect different regions of two or more universes \cite{Wheeler1955, Hawking1988, Li2001, PFGD2003}. Quantum correlations in the form of entanglement among the universal states provide us with a another interaction paradigm in the context of the quantum multiverse and it opens the door to a completely new and wider vision of the multiverse. On the one hand, together with the classical laws of thermodynamics, we can also consider the novel laws of entanglement thermodynamics. This adds a new tool for studying the properties of both the universe and the multiverse. Furthermore, we would expect that the classical and the quantum thermodynamical laws were complementary provided that the quantum theory is a more general framework from which the classical one is recovered as a particular limiting case. Then, local entropic processes of a single universe could be related to the thermodynamical properties of entanglement among universes \cite{RP2012}.

On the other hand, the quantum effects of the space-time are customary restricted to the obscure region of the Planck scale or to the neighbourhood of space-time singularities (both local and cosmological). However, cosmic entanglement among different universes of the multiverse could avoid such restriction and still be present along the whole history of a large parent universe \cite{Mersini2008, RP2011b}. Thus, the effects of inter-universal entanglement on a single universe, and even the boundary conditions of the whole multiverse from which such entanglement would be consequence of, could in principle be tested in a large parent universe like ours. This adds a completely novel feature to the quantum theory of the universe.

The chapter is outlined as follows. In Sec. 2, we shall describe the customary picture in which the universes are spontaneously created from the gravitational vacuum or \emph{space-time foam}. The universes are quantum mechanically described by a wave function that can represent, in the semiclassical regime, either an expanding or a contracting universe. Then, it will be introduced the so-called 'third quantization formalism', where creation and annihilation operators of universes can be defined and it can be given a wave function that represent the quantum state of the multiverse. Afterwards, it will be shown that an appropriate boundary condition of the multiverse allows us to interpret it as made up of entangled pairs of universes.

In Sec. 3, we shall briefly summarize the main features of quantum entanglement in quantum optics, making special emphasis in the characteristics that completely departure from the classical description of light. In Sec. 4, we shall address the question of whether quantum entanglement in the multiverse may induce observable effects in the properties of a single universe. We shall pose a pair of entangled universes and compute the thermodynamical properties of entanglement for each single universe of the entangled pair. It will be shown that the entropy of entanglement can be considered as an arrow of time for single universes and that the vacuum energy of entanglement might allow us to test the whole multiverse proposal. Finally, in Sec. 4, we shall draw some tentative conclusions.

\section{Quantum multiverse}

\subsection{Introduction}

A many-world interpretation of Nature can be dated back to the very ancient Greek philosophy\footnote{The interpretation was posed, of course, in a radically different cultural context. However, it is curious reading some of the pieces that have survived from Greek philosophers like Anaximander, Heraclitus or Democritus, in relation to a 'many-world' interpretation on Nature.} or, in a more recent epoch, to the many-world interpretation that Giordano Bruno derived from the heliocentric theory of Copernicus \cite{Singer1950}, in the XV century, and to the Kant's idea of  'island-universes', term coined by the Prussian naturalist Alenxander von Humboldt in the XIX century \cite{Humboldt1845}. In any case, it was always a very controversial proposal perhaps because the mediocre perception that it entails for our world and for the human being itself.

As it happened historically, the controversy disappears when it is properly defined what it is meant by the word 'world'. If Bruno meant by the word 'world' what is now known as a solar system, von Humbolt meant by 'island universes' what we currently know as galaxies. We now uncontroversially know that there exist many solar systems in billions of different galaxies. Maybe, the  controversy of the current multiverse proposals could partially be unravelled by first  defining precisely what we mean by the word 'universe', in the physics of the XXI century.

Since the advent of the theory of relativity, in the early XX century, we can understand by the word 'universe' a particular geometrical configuration of the space-time as a whole that, following Einstein's equations, is determined by a given distribution of energy-matter in the universe. Furthermore, the geometrical description of the space-time encapsulates the causal relation between material points and, thus, the universe entails everything that may have a causal connection with a particular observer. In other words, the universe is everything we can observe.

Being this true, it does not close the door for the observation of the quantum effects that other universes might have in the properties of our own universe and, thus, it does not prevent us to consider a multiverse scenario. For instance, let us consider a spatially flat space-time endorsed with a cosmological constant. It is well-known that, for a given observer, there is an event horizon beyond which no classical information can be transmitted or received. Thus, two far distant observers are surrounded by their respective event horizons becoming then causally disconnected from each other. These causal enclosures may be interpreted as different universes within the whole space-time manifold\footnote{This is the so-called Level I multiverse in Refs. \cite{Tegmark2003, Tegmark2007}.}. However, cosmic fields are defined upon the whole space-time  and, then, some quantum correlations might be present in the state of the field for two distant regions of the space-time, in the same way as non-local correlations appear in an EPR state of light in quantum optics. Therefore, being two observers classically disconnected, they may share common cosmological quantum fields allowing us, in principle, to study the quantum influence that other regions of the space-time may have in the properties of their isolated patches.

This is an example of a more general kind of multiverse proposals for which it can be defined a common space-time to the universes. It includes the multiverse that comes out in the scenario of eternal inflation \cite{Linde1986, Linde2007}. There are other proposals\footnote{A more exhaustive classification of multiverses and their properties can be found in Refs. \cite{Tegmark2003, Tegmark2007, Mersini2008b}.} in which there is no common space-time among the universes, being the most notable example the landscape of the string theories \cite{Bousso2000, Susskind2003}. In such multidimensional theories, the dimensional reduction that gives rise to our four dimensional universe may contain up to $10^{500}$ different vacua that can be populated with inflationary universes \cite{Hawking2006}. Two universes belonging to different vacua may share no common space-time. However, it might well be that relic quantum correlations may appear between their quantum states, and even some kind of interaction has been proposed to be observable \cite{Mersini2007, Mersini2008}, in principle.

Therefore, even if we have not been exhaustive in the justification of a multiverse scenario, it can easily be envisaged that the multiverse is a plausible cosmological scenario within the framework of the quantum theory provided that this has to be applied to the space-time as a whole. That is  the basic assumption of the present chapter.

\subsection{Classical universes}

In next sections, we shall describe the quantum state of a multiverse made up of homogeneous and isotropic universes. Then, it is worth first noting that homogeneity and isotropy are assumable conditions as far as we deal with large parent universes, where by \emph{large} we mean universes with a length scale which is much greater than the Planck scale even though it can be rather small compared to macroscopic scales. At the Planck length the quantum fluctuations of the metric become of the same order of the metric and the assumptions of homogeneity and isotropy are meaningless. However, except for its very early phase the universe can properly be modeled by a homogeneous and isotropic metric, at least as a first approximation. 

We will also consider homogeneous and isotropic scalar fields. This can be more objectionable. It can be considered a good approximation after the inflationary expansion of the universe has rapidly smoothed out the large inhomogeneities of the distribution of matter in the universe, and it clearly is an appropriate assumption for the large scale of the current universe. However, we should keep in mind that the study of  inhomogeneities is a keystone for the observational tests of the inflationary scenario. Similarly, they might encode valuable information for testing the properties of inter-universal entanglement. However, as a first approach to the problem, we shall mainly be concerned with a multiverse made up of fully homogeneous and isotropic universes and matter fields.

Therefore, let us consider a space-time described by a closed Friedmann-Robertson-Walker (FRW) metric,
\begin{equation}
ds^2 = - \mathcal{N}^2 dt^2 + a^2(t) d\Omega_3^2 ,
\end{equation}
where $\mathcal{N}$ is the lapse function that parameterizes the different foliations of the space-time into space and time, $a(t)$ is the scale factor, and $d\Omega_3^2$ is the usual line element on $S^3$ \cite{Misner1970, Wald1984, Kiefer2007}. The degrees of freedom of the minisuperspace being considered are then the lapse function, $\mathcal{N}$, the scale factor, $a$, and $n$ scalar fields, $\vec{\varphi}=(\varphi_1, \ldots, \varphi_n)$, that represent the matter content of the universe. The total action of the space-time  minimally coupled to the scalar fields can conveniently be written as \cite{Kiefer2007}
\begin{equation}\label{action2}
S = \int dt L = \int dt \mathcal{N} \left( \frac{1}{2} \frac{G_{AB}}{\mathcal{N}^2} \frac{d q^A}{d t} \frac{d q^B}{d t} - \mathcal{V}(q^I)   \right) ,
\end{equation}
for $I, A, B = 0,\ldots, n$, where $G_{AB}\equiv G_{AB}(q^I)$, is the minisupermetric of the $n+1$ dimensional minisuperspace, with $\{ q^I \}\equiv \{a, \vec{\varphi}\}$, and the summation over repeated indices is implicitly understood in Eq. (\ref{action2}). The minisupermetric $G_{AB}$ is given by \cite{Kiefer2007}, $G_{AB} = {\rm diag}(-a, a^3,\ldots, a^3)$, and the potential $\mathcal{V}(q^I)$ by 
\begin{equation}\label{potential}
\mathcal{V}(q^I) \equiv \mathcal{V}(a, \vec{\varphi}) = a^3 \left( V_1(\varphi_1), \ldots, V_n(\varphi_n)\right) - a ,
\end{equation}
where $V_i (\varphi_i)$ is the potential that corresponds to the field $\varphi_i$. The classical equations of motion are obtained by variation of the action (\ref{action2}). Let us for simplicity consider only one scalar field, $\varphi$. Variation of the action with respect to the lapse function, fixing afterwards the value $\mathcal{N}=1$, gives the Friedmann equation
\begin{equation}\label{Fequation}
\left( \frac{d a}{d t} \right)^2 = - 1 + a^2 \sigma^2 \left( \frac{1}{2} \left(\frac{d\varphi}{d t}\right)^2 + V(\varphi)\right) \equiv - 1 + a^2 \sigma^2 \rho_\varphi ,
\end{equation}
where $\rho_\varphi$ is the energy density of the scalar field, and \cite{Linde1986} $\sigma^2 = \frac{8\pi}{3 M_P^2}$, with $M_P \sim 10^{19} {\rm GeV}$ being the Planck mass. Variation of Eq. (\ref{action2}) with respect to the scalar field yields
\begin{equation}\label{Efield}
\frac{d^2\varphi}{d t^2} + \frac{3}{a} \frac{d a}{d t} \frac{d \varphi}{d t} + \frac{\partial V(\varphi)}{\partial \varphi} = 0 .
\end{equation}
Let us focus on a slow-varying scalar field, which constitutes a particularly interesting case that can model the inflationary stage of the universe. In that case \cite{Linde1986, Linde2008}, $\frac{d^2\varphi}{d t^2} \ll  \frac{3}{a} \frac{d a}{d t} \frac{d \varphi}{d t}$ and $(\frac{d \varphi}{d t})^2 \ll V(\varphi)$, and $V(\varphi) \approx V(\varphi_0)$  represents the nearly constant energy density of the scalar field, i.e. $ \rho_\varphi \approx V(\varphi_0)$. 

A limiting case is that of a constant value of the field, $\dot{\varphi}=0$ and  $\rho_\varphi = V(\varphi_0)\equiv \Lambda$. It effectively describes a de-Sitter space-time with a value $\Lambda$ of the cosmological constant. Then, the Friedmann equation (\ref{Fequation}) can be written as
\begin{equation}\label{FLambda}
\frac{d a}{d t} = \sqrt{a^2 H^2 - 1} ,
\end{equation}
where, $H^2 \equiv \sigma^2 \Lambda$. It can be distinguished two regimes. For values, $a\geq \frac{1}{H}$, the real solution 
\begin{equation}\label{aL}
a(t) = \frac{1}{H} \cosh H t ,
\end{equation}
represents a universe that starts out from a value $a_0=\frac{1}{H}$ at $t=0$, and eventually follows an exponential expansion. It corresponds to the Lorentzian regime of the universe. On the other hand, there is no real solution of Eq. (\ref{FLambda}) for values $a< \frac{1}{H}$.  However, we can perform a Wick rotation to Euclidean time, $\tau = i t$, by mean of which Eq. (\ref{FLambda}) transforms into
\begin{equation}\label{FLambdaE}
\frac{d a_E}{d \tau} = \sqrt{1 - a_E^2 H^2} ,
\end{equation}
whose solution,
\begin{equation}\label{aE}
a_E(\tau) = \frac{1}{H} \cos H \tau ,
\end{equation}
is the analytic continuation to Euclidean time of the Lorentzian solution (\ref{aL}). The solution given by Eq. (\ref{aE}) represents an Euclidean space-time that originates at $a_E=0$ (for $\tau = -\frac{\pi}{2H}$), and expands to the value $a_{E}=\frac{1}{H}$ at $\tau = 0$. The transition from the Euclidean region to the Lorentzian region occurs at the boundary hypersurface $\Sigma_0 \equiv \Sigma(a_0)$, at $t=0=\tau$. This transition should not be seen as a process happening \emph{in time} because the Euclidean time is not actual time (it is \emph{imaginary time}). On the contrary, it precisely corresponds to the appearance of time \cite{Kiefer2007} and to the appearance of the (real) universe, actually. 

\begin{figure}[htb]
\begin{minipage}[b]{0.48\linewidth}
\centering
\includegraphics[width=6cm,height=5cm]{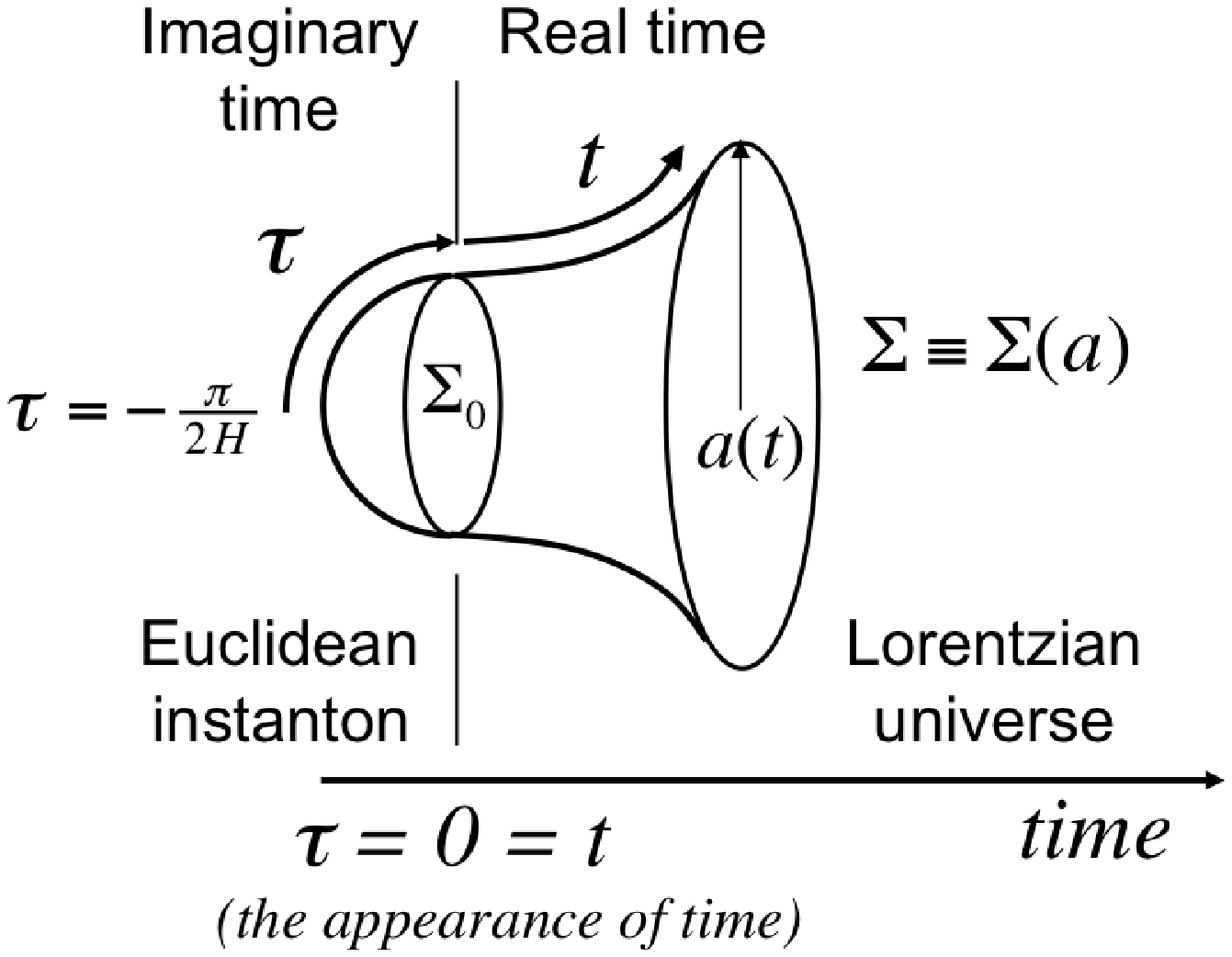}
\caption{The creation of a De-Sitter universe from a De-Sitter instanton.}
\label{DS_instanton}
\end{minipage}
\begin{minipage}[b]{0.50\linewidth}
\vskip-0.5cm
\centering
\includegraphics[width=6.5cm,height=4.55cm]{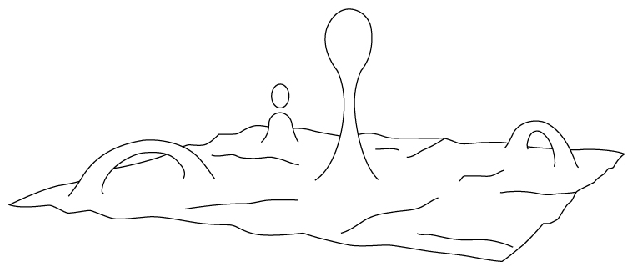} 
\caption{Space-time foam: some of the baby universes may branch off from the parent space-time.}
\label{spacetimefoam}
\end{minipage}
\end{figure}

This is, briefly sketched, the classical picture for the nucleation of a universe from \emph{nothing} \cite{Vilenkin1982, Hawking1983, Kiefer2007}, depicted in Fig. \ref{DS_instanton}, where by \emph{nothing} we should understand a state of the universe where it does not exist space, time and matter, in the customary sense\footnote{However, it does not correspond to the absolute meaning of 'nothing', in a similar way as the vacuum of a quantum field theory is not 'empty' (see, Ref. \cite{Vilenkin1982}).}. Within that picture, the quantum fluctuations of the gravitational vacuum provide it with a \emph{foam structure} \cite{Wheeler1957, Hawking1978, Garay1998, Gott1998} where tiny black holes, wormholes and baby universes \cite{Strominger1990} are virtually created and annihilated (see, Fig. \ref{spacetimefoam}). Some of the baby universes may branch off from the parent space-time and become isolated universes that, subsequently, may undergo an inflationary stage and develop into a large parent universe like ours.

\subsection{Quantum state of the multiverse}

Following the canonical quantization formalism, the momenta conjugated to the configuration variables $q^I$ are given by $p_I \equiv \frac{\delta L}{\delta(\frac{d q^I}{dt})}$, where $L\equiv L(q^I, \frac{d q^I}{dt})$ is the Lagrangian of Eq. (\ref{action2}). The Hamiltonian then reads
\begin{equation}\label{Hamiltonian00}
H \equiv \frac{d q^I}{dt} p_I - L = \mathcal{N} \mathcal{H} \equiv \mathcal{N} \left( G^{AB} p_A p_B + \mathcal{V}(q^I) \right)  .
\end{equation}
The invariance of general relativity under time reparametrizations implies that the variation of the Hamiltonian (\ref{Hamiltonian00}) with respect to the lapse function vanishes. We obtain thus the classical Hamiltonian constraint, $\mathcal{H} = 0$, which gives rise to the Friedmann equation (\ref{Fequation}). The wave function of the universe, $\phi$, can then be obtained by performing a canonical quantization of the momenta, $p_I \rightarrow \hat{p}_I \equiv - i \hbar \frac{\partial}{\partial q^I}$, and applying the quantum version of the Hamiltonian constraint to the wave function $\phi$, i.e. $\hat{\mathcal{H}}\phi = 0$. With an appropriate choice of factor ordering, it can be written as \cite{Kiefer2007}
\begin{equation}\label{WDW0}
\left\{ -\frac{\hbar^2}{\sqrt{-G}} \frac{\partial}{\partial q^A} \left( \sqrt{-G} \, G^{AB} \frac{\partial}{\partial q^B}  \right) + \mathcal{V}(q^I) \right\} \phi(q^I) = 0 ,
\end{equation}
where $G^{AB}$ is the inverse of the minisupermetric $G_{AB}$, with $G^{AB} G_{BC} = \delta^A_C$, and $G$ is the determinant of $G_{AB}$. For a homogeneous and isotropic universe with a slow-varying field the Wheeler-De Witt equation (\ref{WDW0}) explicitly yields
\begin{equation}\label{WDW1}
\hbar^2 \frac{\partial^2 \phi}{\partial a^2} + \frac{\hbar^2}{a} \frac{\partial \phi}{\partial a} + (a^4 V(\varphi)- a^2)\phi = 0 ,
\end{equation}
where, $\phi\equiv \phi(a,\varphi)$. Let us note that if we replace $V(\varphi)$ by $\Lambda$, the wave function $\phi\equiv \phi_\Lambda(a)$ represents the quantum state of a de-Sitter universe. For later convenience, let us write Eq. (\ref{WDW1}) as
\begin{equation}\label{WDW2}
\ddot{\phi} + \frac{\dot{\mathcal{M}}}{\mathcal{M}}  \dot{\phi} + \omega^2 \phi = 0 , 
\end{equation}
where, $\dot{\phi}\equiv \frac{\partial \phi}{\partial a}$ and $\dot{\mathcal{M}}\equiv \frac{\partial \mathcal{M}}{\partial a}$, with $\mathcal{M} \equiv \mathcal{M}(a) = a$, and, $\omega\equiv \omega(a,\varphi) = \frac{a}{\hbar}\sqrt{a^2 V(\varphi)-1}$. It will be useful later on to recall the formal resemblance of Eq. (\ref{WDW2}) to the equation of motion of a harmonic oscillator. The WKB solutions of Eq. (\ref{WDW2}) can be written, in the Lorentzian region, as
\begin{equation}\label{WKBL}
\phi_{WKB}^\pm(a,\varphi) = \frac{N(\varphi)}{\sqrt{\mathcal{M}(a) \omega(a,\varphi)}} e^{\pm i S(a,\varphi)} ,
\end{equation}
where $N(\varphi)$ is a normalization factor, and
\begin{equation}
S(a,\varphi) = \int da \, \omega(a,\varphi) = \frac{1}{\hbar} \frac{(a^2 V(\varphi)- 1)^\frac{3}{2}}{3 V(\varphi)} .
\end{equation}
The positive and negative signs of $\phi_{WKB}^\pm$  correspond to the contracting and expanding branches of the universe, respectively. This can be seen by noticing that, for sufficiently large values of the scale factor, the Fourier transform of $\phi_{WKB}^\pm(a,\varphi)$ is highly peaked around the value of the classical momentum $p_a^c$ \cite{Halliwell1987}, i.e. $\tilde{\phi}_{WKB}^\pm(p_a,\varphi)\approx \delta(p_a - p_a^c)$. The classical momentum reads, $p_a^c = - a \frac{\partial a}{\partial t}$, and quantum mechanically, for large values of the scale factor, $\hat{p}_a \phi = -i\hbar \dot{\phi} \approx \pm \omega \phi$, where the positive and negative signs correspond to the signs of $\phi_{WKB}^\pm$. Then, $\frac{\partial a}{\partial t} \approx \mp \frac{\omega}{a}$, where the negative sign describes a contracting universe and the positive sign an expanding universe. Thus, the solutions $\phi_{WKB}^\pm$ of the Wheeler-de Witt equation (\ref{WDW2}) describe the contracting and expanding branches of the universe, respectively.

\begin{figure}
\includegraphics[width=13cm,height=5cm]{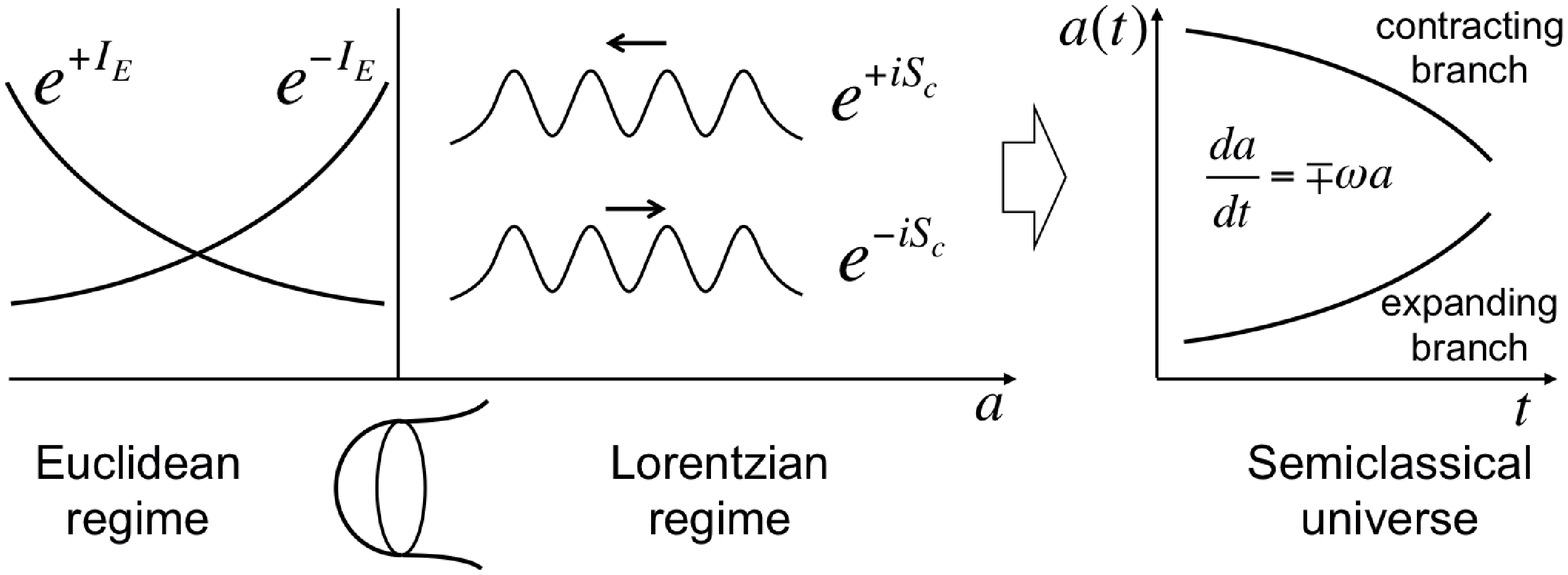}                

\caption{Boundary conditions of the universe. }
\label{ExpCont}
\end{figure}

In order to fix the state of the universe, a boundary condition has to be imposed on the wave function $\phi_{WKB}$. The tunneling boundary condition \cite{Vilenkin1986, Vilenkin1988} states that the only modes that survive the Euclidean barrier are the outgoing modes of the minisuperspace that correspond, in the Lorentzian region, to the expanding branches of the universe (see, Fig. \ref{ExpCont}). Then, the wave function of the universe reads
\begin{equation}\label{TL}
\phi^T(a,\varphi) \approx \frac{N(\varphi)}{\sqrt{\mathcal{M}(a) \omega(a,\varphi)}} e^{- i S(a,\varphi)} ,
\end{equation}
with \cite{Vilenkin1988, Kiefer2007}, $N(\varphi)=e^{-\frac{1}{3 V(\varphi)}}$. By using the matching conditions, the wave function (\ref{TL}) turns out to be given in the Euclidean region by
\begin{equation}\label{TE}
\phi_E^T(a,\varphi) \approx \frac{e^{-\frac{1}{3 V(\varphi)}}}{\sqrt{\mathcal{M}(a) \omega(a,\varphi)}} \left( e^{+ I(a,\varphi)}  + e^{- I(a,\varphi)} \right) ,
\end{equation}
where, $I=iS$, is the Euclidean action. The first term in Eq. (\ref{TE}) may diverge as the scale factor degenerates. However, this is not a problem in terms of Vilenkin's reasoning \cite{Vilenkin1986} because the tunneling boundary condition is mainly intended for fixing the state of the wave function on the Lorentzian region where the current probability is defined, cf. \cite{Vilenkin1986}. The philosophy of the 'no-boundary' proposal of Hartle and Hawking \cite{Hartle1983} is quite the contrary. For these authors, the actual quantum description of the universe is given by a path integral performed over all compact Euclidean metrics. The no-boundary condition is then imposed on the Euclidean sector of the wave function. In the case being considered, it is equivalent to impose regularity conditions \cite{Halliwell1990}, and thus 
\begin{equation}\label{NBE}
\phi_E^{NB}(a,\varphi) \approx \frac{N(\varphi)}{\sqrt{\mathcal{M}(a) \omega(a,\varphi)}}  e^{- I(a,\varphi)}  ,
\end{equation}
with $N(\varphi) = e^{+\frac{1}{3 V(\varphi)}}$ \cite{Halliwell1990, Kiefer2007}. In the Lorentzian sector, the wave function turns out to be given  by a linear combination of expanding and contracting branches of the universe \cite{Hawking1985}, i.e.
\begin{equation}\label{NBL}
\phi^T(a,\varphi) \approx \frac{e^{+\frac{1}{3 V(\varphi)}}}{\sqrt{\mathcal{M}(a) \omega(a,\varphi)}} \cos S  \propto \frac{e^{+\frac{1}{3 V(\varphi)}}}{\sqrt{\mathcal{M}(a) \omega(a,\varphi)}} \left( e^{+ i S(a,\varphi)}  + e^{-i S(a,\varphi)} \right)  .
\end{equation}
Both expanding and contracting branches suffer subsequently a very effective decoherence process \cite{Halliwell1989, Kiefer1992} becoming  quantum mechanically independent. Thus, observers inhabiting a branch of the universe cannot perceive any effect of the quantum superposition.

\subsection{Third quantization formalism}

Let us now introduce the so-called 'third quantization' formalism \cite{Strominger1990}, where the creation and the annihilation of universes is naturally incorporated in a parallel way as the creation and annihilation of particles is naturally formulated in a quantum field theory. The third quantization formalism consists of considering the wave function of the universe, $\phi(a,\vec{\varphi})$, as a field defined upon the minisuperspace of variables $(a,\vec{\varphi})$. The minisupermetric of the minisuperspace, $G_{AB} = {\rm diag}(-a, a^3, \ldots, a^3)$, where $G_{aa}=-a$, has a Lorentzian signature and it allows us to formally interpret the scale factor as an intrinsic time variable of the minisuperspace. This has not to be confused with a time variable in terms of 'clocks and rods' measured by any observer. The consideration of the scale factor as a time variable within a single universe is a tricky task (see Refs. \cite{Vilenkin1986, Isham1992, Hawking1992, Hartle1993, Kiefer1995, Kiefer2005, Kiefer2007}) that will  partially be addressed on subsequent sections.

We already noticed the formal analogy between the Wheeler-de Witt equation (\ref{WDW2}) and the equation of motion of a harmonic oscillator. Taking further the analogy, we can find a (third quantized) action for which the variational principle gives rise to Eq. (\ref{WDW2}), given by
\begin{equation}\label{action3}
^{(3)}S = \frac{1}{2} \int da \,\; ^{(3)}L = \frac{1}{2} \int da \, \left( \mathcal{M} \dot{\phi}^2 - \mathcal{M} \omega^2 \phi^2 \right) .
\end{equation}
The third quantized momentum is defined as, $^{(3)}P_\phi \equiv \frac{\delta \, ^{(3)}L}{\delta \dot{\phi}} = \mathcal{M} \dot{\phi}$, where $^{(3)}L$ is the Lagrangian of the action (\ref{action3}), and the third quantized Hamiltonian then reads
\begin{equation}\label{Hamiltonian3}
^{(3)}H = \frac{1}{2 \mathcal{M}} P_\phi^2 + \frac{\mathcal{M} \omega^2}{2} \phi^2 ,
\end{equation}
where $\mathcal{M}\equiv \mathcal{M}(a)$ and $\omega\equiv \omega(a,\varphi)$ are defined after Eq. (\ref{WDW2}). The configuration variable of the third quantization formalism is the wave function of the universe, $\phi$, and the quantum state of the multiverse is thus given by another wave function, $\Psi \equiv \Psi(\phi, a)$, which is the solution of the (third quantized) Schrödinger equation \cite{Strominger1990, RP2010}
\begin{equation}\label{Schrodinger3}
^{(3)}\hat{H}(\phi, -i \hbar \frac{\partial}{\partial \phi}, a) \Psi(\phi, a) = i\hbar \frac{\partial \Psi(\phi, a)}{\partial a} .
\end{equation}
The customary interpretation of the wave function $\Psi$ is the following \cite{Strominger1990}: let us expand the quantum state of the multiverse, $|\Psi\rangle$, in an orthonormal basis of number states, $|N\rangle$, i.e.
\begin{equation}\label{superposition}
|\Psi\rangle = \sum_N \Psi_N(\phi, a) |N\rangle ,
\end{equation}
then, $|\Psi_N(\phi, a_0)|^2$ gives the probability to find in the multiverse $N$ universes with a value $a_0$ of the scale factor. We can consider different types of universes having different energy-matter contents represented by the fields $\vec{\varphi}^{(i)}$ of the $i$-universe. The wave function of the whole multiverse is given then by a linear superposition of wave functions of the form \cite{RP2010, RP2011b}
\begin{equation}\label{stateMultiverse}
\Psi_{\vec{N}}(\vec{\phi}, a) = \Psi_{N_1}(\phi_1, a)  \Psi_{N_2}(\phi_2, a) \cdots  \Psi_{N_n}(\phi_n, a) ,
\end{equation}
where, $\vec{\phi}\equiv (\phi_1, \phi_2, \ldots, \phi_n)$ and $\vec{N}\equiv (N_1, N_2, \ldots, N_n)$, with $N_i$ being the number of universes of type $i$, represented by the wave function $\phi_i \equiv \phi(\vec{\varphi}^{(i)}, a)$. Following the canonical interpretation of the wave function in quantum mechanics, $|\Psi_{\vec{N}}( \vec{\phi}, a_0)|^2$ gives the probability to find $\vec{N}$ universes in the multiverse with a value of the scale factor and the scalar fields given by, $a=a_0$ and $\vec{\varphi}^{(i)}=\vec{\varphi}_0^{(i)}$, for the $i$-universe.

Let us just consider one type, $i$, of universes. The wave function $\phi_i$ can be promoted to an operator $\hat{\phi}_i$ that can be written as
\begin{equation}
\hat{\phi}_i(a,\varphi) = A_i(a,\varphi) \hat{b}_{0,i}^\dag + A_i^*(a,\varphi) \hat{b}_{0,i} ,
\end{equation}
where the probability amplitudes $A_i(a,\varphi)$ and $A_i^*(a,\varphi)$ satisfy the Wheeler-de Witt equation (\ref{WDW2}), and 
\begin{eqnarray}\label{b01}
\hat{b}_{0,i} \equiv \sqrt{\frac{\mathcal{M}_0 \omega_0}{\hbar}} \left( \hat{\phi}_i + \frac{i}{\mathcal{M}_0 \omega_0} \hat{P}_{\phi_i} \right) , \\ \label{b02}
\hat{b}_{0,i}^\dag \equiv \sqrt{\frac{\mathcal{M}_0 \omega_0}{\hbar}} \left( \hat{\phi}_i - \frac{i}{\mathcal{M}_0 \omega_0} \hat{P}_{\phi_i} \right) ,
\end{eqnarray}
are the customary creation and annihilation operators of the harmonic oscillator, with $\mathcal{M}_0$ and $\omega_0$ being the mass and frequency terms, $\mathcal{M}(a)$ and $\omega(a,\varphi)$, respectively, evaluated on the boundary hypersurface $\Sigma_0$ for which, $a=a_0$ and $\varphi=\varphi_0$. The operators $\hat{b}_{0,i}$ and $\hat{b}_{0,i}^\dag$ can then be interpreted as the annihilation and creation operators of universes with a value of the scale factor $a_0$ and an energy density given by $\rho_\varphi\approx V(\varphi_0)$, for the case of a slow-varying field. The kind of universes created and annihilated by $\hat{b}_{0,i}^\dag$ and $\hat{b}_{0,i}$, respectively, also depend on the boundary conditions imposed on the probability amplitudes $A_i(a,\varphi)$ and $A_i^*(a,\varphi)$. Recalling the previous discussion on the boundary conditions of the universe, if the tunneling boundary condition is imposed, then, $\hat{b}_{0,i}^\dag$ ($\hat{b}_{0,i}$) creates (annihilates) expanding branches of the universe. If otherwise the 'no-boundary' proposal is chosen, $\hat{b}_{0,i}^\dag$ ($\hat{b}_{0,i}$) creates (annihilates) linear combinations of expanding and contracting branches.

Therefore, at least for universes with high order of symmetry, the third quantization formalism parallels that of a quantum field theory in a curved space-time, i.e. it can formally be seen as a quantum field theory defined on the curved minisuperspace described by the minisupermetric $G_{AB}$. The scale factor formally plays the role of the time variable and the matter fields $\vec{\varphi}$ the role of the spatial coordinates. Creation and annihilation operators of universes can properly be defined in the curved minisuperspace. However, as it happens in a quantum field theory, different representations can be chosen to describe the quantum state of the universes. The meaning of such representations needs of a further analysis in terms of the boundary condition that has to be imposed on the quantum state of the whole multiverse.

\subsection{Boundary conditions of the multiverse}

For a given representation, $\hat{b}_i^\dag$ and $\hat{b}_i$, the eigenvalues of the number operator $\hat{N}_i \equiv \hat{b}_i^\dag\hat{b}_i$ might be interpreted in the third quantization formalism as the number of $i$-universes in the multiverse, where the index $i$ labels the different kinds of universes considered in the model. However, in terms of the constant operators $\hat{b}_{0,i}$ and $\hat{b}_{0,i}^\dag$ defined in Eqs. (\ref{b01}-\ref{b02}), the number of universes of the multiverse is not conserved because $\hat{N}_{0,i}\equiv \hat{b}_{0,i}^\dag \hat{b}_{0,i}$ is not an invariant operator, i.e.
\begin{equation}
\frac{d \hat{N}_{0,i}}{d a} \equiv \frac{i}{\hbar} [^{(3)}\hat{H}_i, \hat{N}_{0,i}] + \frac{\partial \hat{N}_{0,i}}{\partial a} = \frac{i}{\hbar} [^{(3)}\hat{H}_i, \hat{N}_{0,i}] \neq 0 .
\end{equation}
For a large parent universe, i.e. for values $a\gg1$, the creation and annihilation operators can asymptotically be taken to be the usual creation and annihilation operators of the harmonic oscillator (\ref{Hamiltonian3}) with the proper frequency $\omega$ of the Hamiltonian, i.e.
\begin{eqnarray}\label{properRepresentation1}
\hat{b}_{\omega,i} \equiv \sqrt{\frac{\mathcal{M}(a) \omega_i(a,\varphi)}{\hbar}} \left( \hat{\phi}_i + \frac{i}{\mathcal{M}(a) \omega_i(a,\varphi)} \hat{P}_{\phi_i} \right) , \\ \label{properRepresentation2}
\hat{b}_{\omega,i}^\dag \equiv \sqrt{\frac{\mathcal{M}(a) \omega_i(a,\varphi)}{\hbar}} \left( \hat{\phi}_i - \frac{i}{\mathcal{M}(a) \omega_i(a,\varphi)} \hat{P}_{\phi_i} \right) ,
\end{eqnarray}
for a given type of $i$-universes. However, in terms of the asymptotic representation (\ref{properRepresentation1}-\ref{properRepresentation2}) the number operator, $\hat{N}_{\omega,i} \equiv \hat{b}_{\omega,i}^\dag \hat{b}_{\omega,i}$, is neither an invariant operator because
\begin{equation}
\frac{d \hat{N}_{\omega,i}}{d a} \equiv \frac{i}{\hbar} [^{(3)}\hat{H}_i, \hat{N}_{\omega,i}] + \frac{\partial \hat{N}_{\omega,i}}{\partial a} = \frac{\partial \hat{N}_{\omega,i}}{\partial a}  \neq 0 .
\end{equation}
It would be expected that the number of universes in the multiverse would be a property of the multiverse independent of any internal property of a particular single universe. Therefore, it seems appropriate to impose the following boundary condition on the multiverse:
\begin{center}
\fbox{\parbox{12cm}{The number of universes of the multiverse does not depend on the value of the scale factor of a particular single universe.}}
\end{center}
This boundary condition imposes the restriction that the number operator $\hat{N}_i$ for a particular type of $i$-universes has to be an invariant operator\footnote{We are not considering transitions from one kind of universes to another.}. We can then follow the theory of invariants developed by Lewis \cite{Lewis1969} and others \cite{Pedrosa1987, Dantas1992, Sheng1995, Song2000, Kim2001, Park2004, Vergel2009}, and find a Hermitian invariant operator, $\hat{I}_i= \hbar (\hat{b}_i^\dag \hat{b}_i + \frac{1}{2})$, where \cite{Lewis1969}
\begin{eqnarray}\label{b1}
\hat{b}_i(a) &\equiv& \sqrt{\frac{1}{2 \hbar}} \left( \frac{1}{R_i} \hat{\phi}_i + i (R_i \hat{P}_{\phi_i} - \dot{R}_i \hat{\phi}_i) \right) , \\ \label{b2}
\hat{b}_i^\dag(a) &\equiv& \sqrt{\frac{1}{2 \hbar}} \left( \frac{1}{R_i} \hat{\phi}_i  - i (R_i \hat{P}_{\phi_i} - \dot{R}_i \hat{\phi}_i) \right) , 
\end{eqnarray}
with, $R_i\equiv R_i(a,\varphi)$, that can be written as $R=\sqrt{\phi_{1,i}^2 + \phi_{2,i}^2}$, being $\phi_{1,i}$ and $\phi_{2,i}$ two independent solutions of the Wheeler-de Witt equation (\ref{WDW2}). In the semiclassical regime, we can use independent combinations of the solutions $\phi^{WKB}_+$ and $\phi^{WKB}_-$ so that
\begin{equation}\label{R}
R_i(a,\varphi) \approx \frac{e^{\pm\frac{1}{3V_i(\varphi)}}}{\sqrt{\mathcal{M}(a) \omega_i(a,\varphi)}} ,
\end{equation}
where the positive sign corresponding to the choice of the no-boundary proposal and the negative sign to the tunneling boundary condition. The number operator for a particular kind of $i$-universes in the representation given by Eqs. (\ref{b1}-\ref{b2}), $\hat{N}_i\equiv \hat{b}_i^\dag \hat{b}_i$, is then an invariant operator fulfilling the boundary condition of the multiverse and, thus, the eigenvalues $N_i$, with $\hat{N}_i |N_i, a\rangle = N_i |N_i, a\rangle$ and $N_i\neq N_i(a)$, can properly be interpreted as the number of $i$-universes of the multiverse. 

In terms of the invariant representation, the Hamiltonian (\ref{Hamiltonian3}) takes the form
\begin{equation}\label{Hamiltonian4}
^{(3)}\hat{H} = \hbar \left(  \beta_+ \,  (\hat{b}^\dag)^2 + \beta_-  \, \hat{b}^2 + \beta_0  \, (\hat{b}^\dag \hat{b} + \frac{1}{2})  \right) ,
\end{equation}
where,
\begin{eqnarray}\label{beta+}
\beta_+^* = \beta_- &=& \frac{1}{4} \left\{ \left( \dot{R} - \frac{i}{R} \right)^2 + \omega^2 R^2  \right\} , \\ \label{beta0}
\beta_0 &=& \frac{1}{2} \left( \dot{R}^2 + \frac{1}{R^2} + \omega^2 R^2 \right) .
\end{eqnarray}
The Hamiltonian (\ref{Hamiltonian4}) is formally the same Hamiltonian of a degenerated parametric amplifier used in quantum optics \cite{Scully1997, Walls2008} (see also, Sec. 3). The quadratic terms are interpreted therein as the creation and annihilation operators of pairs of entangled photons. Similarly, we can interpret the quadratic terms in $\hat{b}^\dag$ and $\hat{b}$ of Eq. (\ref{Hamiltonian4}) as operators that create and annihilate, respectively, pairs of entangled universes. In the case that the universes were distinguishable, the Hamiltonian (\ref{Hamiltonian4}) would take the form of a non-degenerated parametric amplifier \cite{Walls2008}
\begin{equation}\label{Hamiltonian4b}
^{(3)}\hat{H} = \hbar \left(  \beta_+  \, \hat{b}_1^\dag \hat{b}_2^\dag + \beta_-  \, \hat{b}_1 \hat{b}_2 + \frac{\beta_0}{2} \,  (\hat{b}_1^\dag \hat{b}_1 + \hat{b}_2^\dag \hat{b}_2 + 1)  \right) ,
\end{equation}
where the indices $1$ and $2$ label the two universes of the entangled pair. The distinguishability of universes is certainly a tricky task. However, observers may exist in the two universes of an entangled pair because the universes share similar properties and, then, the plausible (classical and quantum) communications between these observers would make the universes be distinguishable. Classical communications between the observers of different universes can be conceivable by the presence of wormholes connecting the universes and quantum communications could then be implemented by using quantum correlated fields shared by the two observers. Therefore, it is at least plausible to pose a model of the multiverse made up of entangled pairs of distinguishable universes.

The general quantum state of a multiverse formed by entangled pairs of de-Sitter universes would  be given by linear combinations of terms like \cite{RP2010, RP2011b} (see Eq. (\ref{stateMultiverse}))
\begin{equation}\label{stateMultiverse2}
\Psi_{\vec{N}}(\vec{\phi}, a) = \Psi_{N_1}^{\Lambda_1}(a, \phi_1)  \Psi_{N_2}^{\Lambda_2}(a, \phi_2) \cdots  \Psi_{N_n}^{\Lambda_n}(a, \phi_n) ,
\end{equation}
where, $\vec{\phi}\equiv (\phi_1, \phi_2, \ldots, \phi_n)$, and $\vec{N}\equiv (2 N_1, 2 N_2, \ldots, 2 N_n)$, with $N_i$ being the number of pairs of universes of type $i$, represented by the wave function $\phi_i \equiv \phi_{\Lambda_i}(a)$ that corresponds to the value $\Lambda_i$ of the cosmological constant. The wave functions, $\Psi_{N_i}^{\Lambda_i}(\phi_i, a)$, in Eq. (\ref{stateMultiverse2}) are the solutions of the third quantized Schr\"{o}dinger equation
\begin{equation}\label{3Schrodinger}
i \hbar \frac{\partial}{\partial a} \Psi_{N_i}^{\Lambda_i}(\phi_i, a) = \hat{H}_i(\phi, p_\phi, a) \Psi_{N_i}^{\Lambda_i}(\phi_i, a) ,
\end{equation}
with
\begin{equation}\label{HamiltonianTotalMultiverse}
\hat{H}_i = \hbar \left\{ \beta_{-}^{(i)} \hat{b}_{1}^{(i)} \hat{b}_{2}^{(i)} + \beta_{+}^{(i)} (\hat{b}_{1}^{(i)})^\dag (\hat{b}_{2}^{(i)})^\dag  + \frac{1}{2} \beta_{0}^{(i)} \left( (\hat{b}_{1}^{(i)})^\dag \hat{b}_{1}^{(i)}  + (\hat{b}_{2}^{(i)})^\dag \hat{b}_{2}^{(i)} + 1\right)  \right\} ,
\end{equation}
for each kind of $i$-universes in the multiverse \cite{RP2011b}.

\section{Quantum entanglement}

\subsection{Introduction}

Back to the early years of the quantum development, in 1935, Schrödinger \cite{Schrodinger1936b, Schrodinger1936a} coined the word 'entanglement' to describe a puzzling feature of the quantum theory that was formerly posed by Einstein, Podolski and Rosen in a famous gedanken experiment \cite{Einstein1935}. Schrödinger also realized that entanglement is precisely \emph{the characteristic trait of quantum mechanics, the one that enforces its entire departure from classical lines of thought} \cite{Schrodinger1936b}.  Let us briefly show it by following the example given in Ref. \cite{Scully1997} (see also Ref. \cite{Walls2008}). Let us consider the photo-disintegration of a Hg$_2$ molecule formed by two atoms of Hg with spin $\frac{1}{2}$. Before the disintegration, the molecule is taken to be in a state of zero angular momentum so that the composite state is given by
\begin{equation}\label{es00}
|{\rm Hg}_2\rangle = \frac{1}{\sqrt{2}} \left( | \uparrow_1 \downarrow_2 \rangle - | \downarrow_1 \uparrow_2 \rangle \right) , 
\end{equation}
where $1$ and $2$ refer to the atoms of Hg and $|\uparrow(\downarrow) \rangle$ refers to the value $+\frac{1}{2}(-\frac{1}{2})$ of the projection of their spin along the $z$-axis. After the photo-disintegration, performed with no disturbance of the angular momentum, the two atoms separate each other in opposite directions so we can make independent measurements on them. Before doing any measurement we do not know the particular value of the spin of each atom. However, we do anticipatedly know that if a measurement of the spin projection is performed on the atom $1$ yielding a value $+\frac{1}{2}(-\frac{1}{2})$, then, the spin projection of the atom $2$ is to be $-\frac{1}{2}(+\frac{1}{2})$. Furthermore, if it is performed a different measurement of the projection of the spin of the particle $1$ along, say, the $x$-axis, we are determining the value of the spin projection of the particle $2$ along the same axis, too. This non-local feature of the quantum theory is known as \emph{entanglement}  and the state (\ref{es00}) is called an \emph{entangled state}. 

In 1964, Bell derived certain inequalities \cite{Bell1964, Bell1987} that should be satisfied by any reasonable realistic\footnote{By a realistic theory we mean a theory that presupposes that the elements of the theory represent elements of physical reality (see Ref. \cite{Einstein1935}).} theory of local variables. The experiments of Aspect \cite{Aspect1982} and others \cite{Tittel1998, Weihs1998, Rowe2001, Ansmann2009} have shown that the entangled states of the quantum theory violate such inequalities. Furthermore, these  states have not only provided us with an experimental test of the quantum postulates but they have also given rise to the development of a completely new branch of physics, the so-called quantum information theory \cite{Vedral2006, Jaeger2007, Horodecki2009}, which includes interesting subjects like quantum computation, quantum cryptography, and quantum teleportation, which are currently under a promising state of development.

It is finally worth noticing that the kinematical non-locality of the quantum theory is also the feature that forces us to consider a wave function of the universe. As it is pointed out in Ref. \cite{Kiefer2007}, \emph{if gravity is quantized, the kinematical non-separability of quantum theory demands that the whole universe must be described in quantum terms} (cf. p. 4). Every space-time region is entangled to its environment, which is entangled to another environment and so forth, ending up in a quantum description of the whole universe.

\subsection{Squeezed and entangled states of light}

Squeezed states of light \cite{Walls1983} can be seen as a generalization of the coherent states. Let  us define the quadrature operators
\begin{equation}\label{quadrature}
\hat{X}_1 \equiv \hat{a} + \hat{a}^\dag \;\; , \;\; \hat{X}_2 = i (\hat{a}^\dag - \hat{a}) ,
\end{equation}
where $\hat{a}^\dag$ and $\hat{a}$ are the usual creation and annihilation operators of the harmonic oscillator. The operators $\hat{X}_1$ and $\hat{X}_2$ are essentially dimensionless position and momentum operators. The uncertainty relation for $\Delta X_1$ and $\Delta X_2$ reads
\begin{equation}
\Delta X_1 \Delta X_2 \geq 1 ,
\end{equation}
where, for a coherent state, $\Delta X_1 = \Delta X_2 = 1$. A squeezed state is defined as the quantum state for which one of the quadratures  satisfies\footnote{\label{fn01}An ideal squeezed state also satisfies $\Delta X_1 \Delta X_2 = 1$}
\begin{equation}
(\Delta X_i)^2 < 1 \; \; (i = 1 \; {\rm or} \; 2) .
\end{equation}
Therefore, for a squeezed state the uncertainty of one of the quadratures is reduced below the limit of the Heisenberg principle at the expense of the increased fluctuations of the other quadrature. 

Unlike the generation of coherent states, which is associated with linear terms of the creation and annihilation operators in the Hamiltonian, the generation of squeezed states is associated with quadratic terms of such operators. For instance, let us consider the Hamiltonian that represents in quantum optics a degenerated parametric amplifier \cite{Scully1997, Walls2008}
\begin{equation}\label{degeneratedAmplifier}
\hat{H} = i \hbar \frac{\chi}{2}  \left( (\hat{a}^\dag)^2 - \hat{a}^2 \right) ,
\end{equation}
where $\chi$ is a coupling constant. Then, the time evolution of the vacuum state,
\begin{equation}
|s(t)\rangle = \hat{S}(\chi) |0\rangle = e^{\frac{\chi}{2}\left( (\hat{a}^\dag)^2 - \hat{a}^2 \right) t} |0\rangle ,
\end{equation}
yields a squeezed (vacuum) state, $|s(t)\rangle$, with $\hat{S}_\chi$ being the squeezing operator which satisfies, $\hat{S}^\dag(\chi) = \hat{S}^{-1}(\chi) = \hat{S}(-\chi)$. It is therefore a unitary operator. The Heisenberg equations of motion for the quadrature amplitudes turn out to be then
\begin{equation}
\frac{d \hat{X}_1}{d t} = \chi \hat{X}_1 \; \; , \; \; \frac{d \hat{X}_2}{d t} = - \chi \hat{X}_2 ,
\end{equation}
with solutions given by
\begin{equation}
\hat{X}_1(t) = e^{\chi t} \hat{X}_1(0) \; \; , \; \; \hat{X}_2(t) = e^{-\chi t} \hat{X}_2(0) .
\end{equation}
Then, for an initial vacuum state, for which $\Delta X_i(0) = 1$, the variances of the quadratures read
\begin{equation}
\Delta X_1(t) = e^{2\chi t} \; \; , \; \; \Delta X_2(t) = e^{- 2 \chi t} .
\end{equation}
It can clearly be seen that one of the variances ($\Delta X_2$) decreases in time at the expense of the increase of the other ($\Delta X_1$), with $\Delta X_1(t) \Delta X_2(t) = 1$. The squeezed vacuum state is therefore an ideal squeezed state (see footnote \ref{fn01}).

The Hamiltonian given by Eq. (\ref{degeneratedAmplifier}) is associated with the generation of entangled pairs of photons of equal frequency. For that reason, squeezed states are usually dubbed two photon coherent states \cite{Yuen1975, Yuen1976}. The non-degenerate amplifier is a generalization of the Hamiltonian (\ref{degeneratedAmplifier}) which generates entangled pairs of distinguishable photons of frequency $\omega_1$ and $\omega_2$, respectively. In that case, the Hamiltonian reads
\begin{equation}\label{ngpa}
\hat{H} = i \hbar \chi (\hat{a}_1^\dag \hat{a}_2^\dag - \hat{a}_1 \hat{a}_2 ) ,
\end{equation}
where $\hat{a}_1^\dag$, $\hat{a}_1$ and $\hat{a}_2^\dag$, $\hat{a}_2$ are the creation and annihilation operators of modes with frequency $\omega_1$ and $\omega_2$, respectively. The solutions of the Heisenberg equations read \cite{Walls2008}
\begin{eqnarray}\label{so01}
\hat{a}_1(t) &=& \hat{a}_1(0) \cosh\chi t + \hat{a}_2^\dag(0) \sinh\chi t , \\ \label{so02}
\hat{a}_2(t) &=& \hat{a}_2(0) \cosh\chi t + \hat{a}_1^\dag(0) \sinh\chi t ,
\end{eqnarray}
and the evolution of the two-mode vacuum state is now given by
\begin{equation}\label{twomodeS}
|s_2 \rangle = \hat{S}_2(\chi) |0_1 0_2\rangle = e^{\left( \hat{a}_1^\dag \hat{a}_2^\dag - \hat{a}_1 \hat{a}_2 \right) \chi t} |0_1 0_2 \rangle ,
\end{equation}
where $\hat{S}_2(\chi)$ is the two mode squeeze operator.

\begin{figure}[htb]
\centering
\includegraphics[width=5.3cm,height=4.4cm]{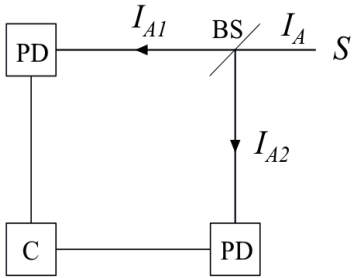}  
\caption{Experimental setup for testing photon antibunching \cite{Reid1986}: S, source of light; BS, beam splitter; PD, photodetector; and, C, correlator.}
\label{BS}
\end{figure}

Squeezed and entangled states are usually dubbed non-classical states \cite{Reid1986} because they may violate some inequalities that should be satisfied in the classical description of light. For instance, in Fig. \ref{BS} it is depicted the typical experimental setup to test the violation of the classical inequality $g^{(2)}(0) \geq 1$ (photon bunching \cite{Reid1986, Walls2008}), where $g^{(2)}(\tau)$ is the second order correlation function that measures the correlation between the state of the field at two different times $t$ and $t+\tau$. Classically, a beam of light with an initial intensity $I_A$ is split into two beams of equal intensities, $I_{A1} = I_{A2} \equiv I$. If the averaged  intensity is defined by
\begin{equation}
\langle I \rangle = \int P(I) I \, dI ,
\end{equation}
for a given positive distribution $P(I)$, then, $g^{(2)}(0)$ can be written as
\begin{equation}\label{ccorrelation}
g^{(2)}(0) = \frac{\langle I_{A1} I_{A2} \rangle}{\langle I_{A1} \rangle \langle I_{A2} \rangle} = \frac{\langle I^2 \rangle}{\langle I \rangle^2} = 1 + \frac{1}{\langle I\rangle^2} \int dI \, P(I) (I - \langle I \rangle )^2  \geq 1 .
\end{equation}
Quantum mechanically, however, the second order correlation function is defined, for a single mode, as \cite{Reid1986, Walls2008}
\begin{equation}\label{qcorrelation}
g^{(2)}(0) = \frac{\langle (a^\dag)^2 a^2 \rangle}{\langle a^\dag a \rangle^2} \geq 1 - \frac{1}{\langle a^\dag a \rangle}  .
\end{equation}
There is then room for a quantum violation of the classical inequality $g^{(2)}(0) \geq 1$. For a large number of photons the quantum inequality (\ref{qcorrelation}) becomes the classical constraint $g^{(2)}(0) \geq 1$, and light can be described classically.

It is worth noticing that what it is violated in an experimental setup involving squeezed and entangled states are some classical assumptions. For instance, in the experimental setup depicted in Fig. \ref{BS}, the photon is not split into two photons by the beam splitter but it takes either the path that reaches the photo-counter $1$ or the path that reaches the photo-counter $2$. The fact that the photon is not divided into two photons, as it would happen to an electromagnetic wave, supports the consideration of the photon as a real and individual entity. Moreover, the corpuscular nature of the photon is the postulate that Einstein assumed in order to properly describe the photoelectric effect and it can be considered the germ of quantum mechanics, actually.

However, such a conclusion does not imply that we can interpret the photon as a classical particle. The double-slit experiment clearly shows that the concept of photon as a localized particle is generally meaningless. The quantum concept of particle has rather to be understood as a global property of the field. Their localization and the space-time  independence of different particles depend on the \emph{separability} of their states. Furthermore, the violation of the classical inequalities is associated with negative values of the probability distributions. This can clearly be seen from Eq. (\ref{ccorrelation}), where a negative value of $P(I)$ is needed to obtain a value $g^{(2)}(0) < 1$. It plainly shows that there are quantum states of light that cannot be described classically \cite{Reid1986}.

Another test for the non-classicality of some quantum states is given by the violation of the Bell's inequalities. This is achieved, for a two mode state of light, whenever it is satisfied \cite{Reid1986}
\begin{equation}
C \equiv \frac{\langle a_1^\dag a_1 a_2^\dag a_2 \rangle}{\langle a_1^\dag a_1 a_2^\dag a_2 \rangle + \langle (a_1^\dag)^2 a_1^2 \rangle} \geq \frac{\sqrt{2}}{2} .
\end{equation}
For the two mode squeezed operators (\ref{so01}-\ref{so02}), it can be checked that
\begin{eqnarray}
\langle a_1^\dag a_1 a_2^\dag a_2 \rangle &=& N^2 (6 x^4 + 6 x^2 + 1) + N (6 x^4 + 4x^2) + x^2 (2 x^2 + 1) , \\
\langle (a_1^\dag)^2 a_1^2 \rangle &=& N^2 (6 x^4 + 6 x^2 + 1) + N (6 x^4 + 2 x^2 - 1) + 2 x^4 ,
\end{eqnarray}
where, $x \equiv x(t) = \sinh \chi t$, and the mean value has been computed for initial number states, with $N_1 = N_2 \equiv N$. For an initial vacuum state, $x(0) = 0$ and $N=0$, then $C=1>0.7$, which implies a maximum violation of Bell's inequalities\footnote{Let us notice that for a pure entangled state like $|\psi\rangle = \frac{1}{\sqrt{2}} (|00\rangle + | 11\rangle)$,  $\langle \psi | (a_1^\dag)^2 a_1^2|\psi \rangle = 0$ and thus $C=1$, too.}. This result is expected because the quantum vacuum state is a highly non-local state. For a pair of entangled photons ($N=1$), it is obtained 
\begin{equation}
C = \frac{14 x^4 + 11 x^2 + 1}{28 x^4 + 19 x^2 + 1} ,
\end{equation}
which implies a violation of Bell's inequalities for a value, $0.31 > \sinh\chi t > 0$. At later times, the effective number of photons, $\langle N_{eff}\rangle = \sinh^2\chi t$, produced by the parametric amplifier grows and the quantum correlations are destroyed. The radiation effectively becomes classical, then. However, at shorter times, the two mode squeezed states violate the Bell's inequalities showing their non-classical behaviour.

Therefore, entangled and squeezed states can essentially be seen as  non-classical states, which is fundamentally related to the complementary principle of quantum mechanics. Generally speaking, the classical description of light in terms of waves and particles, separately, does not hold: i) the photon has to be considered as an individual entity (particle description), and ii) we have to complementary consider interference as well as non-local effects between the states of two distant photons (wave description).

\subsection{Thermodynamics of entanglement}

For a physical system whose quantum state is represented by a density matrix\footnote{In this section, it turns out to be convenient to use the density matrix formalism. This can generally be found in the bibliography (see, for instance, Refs. \cite{Neumann1996, Vedral2006, Jaeger2007, Horodecki2009, Gemmer2009}). Let us just briefly note that for a pure state $|\psi\rangle$, the density matrix is given by $\hat{\rho} = |\psi\rangle \langle \psi |$, and for a mixed state, $\hat{\rho} = \sum_i \lambda_i | i \rangle \langle i |$, where $\lambda_i < 1$ are the eigenvalues of the density matrix, with $\sum_i \lambda_i = 1$, and the vectors $\{ | i \rangle\}$ form an orthonormal basis.}, $\hat{\rho}(t)$, whose evolution is determined by a Hamiltonian, $\hat{H}\equiv \hat{H}(t)$, we can define the following thermodynamical quantities \cite{Alicki2004, Gemmer2009}
\begin{eqnarray}\label{QE}
E(t) &=& {\rm Tr} ( \hat{\rho}(t) \hat{H}(t) ) , \\ \label{QQ}
Q(t) &=& \int^t {\rm Tr} \left( \frac{d \hat{\rho}(t')}{dt'} \hat{H}(t') \right) dt' , \\ \label{QW}
W(t) &=& \int^t {\rm Tr} \left( \hat{\rho}(t') \frac{d\hat{H}(t')}{dt'} \right) dt' ,
\end{eqnarray}
where ${\rm Tr}(\hat{O})$ denotes the trace of the operator $\hat{O}$. The quantities $E(t)$, $Q(t)$, and $W(t)$, are the quantum informational analogue to the energy, heat and work, respectively. The first principle of thermodynamics,
\begin{equation}
d E = \delta W + \delta Q ,
\end{equation}
is then directly satisfied. The quantum informational analogue to the entropy is defined through the von Neumann formulae \cite{Neumann1996, Vedral2006, Jaeger2007, Gemmer2009}
\begin{equation}\label{QS}
S(\rho) = - {\rm Tr} \left( \hat{\rho}(t) \ln \hat{\rho}(t) \right) = - \Sigma_i \lambda_i(t) \ln \lambda_i(t) ,
\end{equation}
where $\lambda_i(t)$ are the eigenvalues of the density matrix, and $0 \ln 0 \equiv 0$. For a pure state, $\hat{\rho}^n = \hat{\rho}$ and $\lambda_i = \delta_{ij}$ for some value $j$. Then, the entropy vanishes. For a mixed state, $S > 0$. It can be distinguished two terms  \cite{Alicki2004} in the variation of entropy,
\begin{equation}\label{variationS}
d S = \frac{\delta Q}{T} + \sigma .
\end{equation}
The first term corresponds to the variation of the entropy due to the change of heat. The second term in Eq. (\ref{variationS}) is called \cite{Alicki2004} \emph{entropy production}, and it accounts for the variation of entropy due to any adiabatic process. The second principle of thermodynamics states that  the change of entropy has to be non-negative for any adiabatic process, i.e. $\sigma \geq 0$.

Let us now analyze the thermodynamical properties of a two mode squeezed state, Eq.  (\ref{twomodeS}), represented by the density matrix
\begin{equation}
\hat{\rho} = |s_2\rangle \langle s_2| = \hat{S}_2(r) |0_1 0_2 \rangle \langle 0_1 0_2 | \hat{S}_2^\dag(r) ,
\end{equation}
where the squeezing operator is given by, $\hat{S}_2(r) \equiv e^{(\hat{a}_1^\dag \hat{a}_2^\dag - \hat{a}_1 \hat{a}_2) r(t)}$, with $r(t) = \chi t$, and $|0_1 0_2\rangle \equiv |0_1\rangle |0_2\rangle$, with $|0_1\rangle$ and $|0_2\rangle$ being the initial ground states of each single mode, respectively. The reduced density matrix that represents the quantum state of each single mode can be obtained by tracing out the degrees of freedom of the partner mode, i.e.
\begin{equation}\label{reduced00}
\hat{\rho}_1 \equiv {\rm Tr}_2 \hat{\rho} = \sum_{N_2=0}^\infty \langle N_2 | \hat{\rho} \ N_2 \rangle ,
\end{equation}
and similarly for $\hat{\rho}_2$ by replacing the indices $2$ and $1$. By making use of the disentangling theorem \cite{Wodkiewicz1985, Buzek1989}, the squeezing operator $\hat{S}_2(r)$ can be written as
\begin{equation}
\hat{S}_2(r) = e^{\Gamma(t) \hat{a}_1^\dag \hat{a}_2^\dag} e^{- g(t) (\hat{a}_1^\dag \hat{a}_1 + \hat{a}_2^\dag \hat{a}_2 + 1)} e^{-\Gamma(t) \hat{a}_1 \hat{a}_2} ,
\end{equation}
where
\begin{equation}
\Gamma(t) \equiv \tanh r(t) \; \; , \; \; g(t) \equiv \ln \cosh r(t) ,
\end{equation}
with, $r(t) = \chi t$. We  can thus compute the reduced density matrix (\ref{reduced00}), yielding
\begin{equation}
\hat{\rho}_1(t) = e^{- 2 g(t)} \sum_{N_1=0}^\infty e^{2 N_1 \ln \Gamma(t)} |N_1\rangle \langle N_1 | = \frac{1}{\cosh^2r(t)} \sum_{N_1=0}^\infty \left( \tanh^2r(t) \right)^{N_1} |N_1\rangle \langle N_1 | .
\end{equation}
It turns out to be that $\hat{\rho}_1$ describes a thermal state 
\begin{equation}\label{thermal00}
\hat{\rho}_1(t) = \frac{1}{Z(t)} \sum_{N_1=0}^\infty e^{-\frac{\omega_1}{T(t)} (N_1 + \frac{1}{2})} |N_1\rangle \langle N_1 | ,
\end{equation}
where, $Z^{-1} = 2 \sinh\frac{\omega_1}{2 T(t)}$, with a time dependent temperature of entanglement given by
\begin{equation}
T(t) = \frac{\omega_1}{2 \ln\frac{1}{\Gamma(t)}} ,
\end{equation}
with $\omega_1$ being the frequency of the mode. It is worth mentioning that the thermal state (\ref{thermal00}) is indistinguishable from a classical mixture \cite{Partovi2008, Jennings2010}. In that sense, it can be seen as a classical state. However, it has been obtained from the  partial trace of a composite entangled state which is, as it has  previously been shown, a quantum state having no classical analogue.

We can now compute the thermodynamical quantities given by Eqs. (\ref{QE}-\ref{QW}) and Eq. (\ref{QS}) for the thermal state (\ref{thermal00}). The entropy of entanglement, i.e. the quantum entropy that corresponds to the reduced density matrix $\hat{\rho}_1$, reads
\begin{equation}\label{entropy00}
S_{ent}(t) = - {\rm Tr}(\hat{\rho}_1 \ln \hat{\rho}_1) = \cosh^2 r(t) \ln\cosh^2r(t) - \sinh^2r(t) \ln\sinh^2r(t) .
\end{equation}
The total energy $E_1\equiv E(\rho_1)$ yields
\begin{equation}
E_1(t) = \omega_1 (\sinh^2 r(t) + \frac{1}{2}) \equiv \omega_1 (\langle N(t) \rangle + \frac{1}{2}) ,
\end{equation}
where $\langle N(t) \rangle$ is an effective mean number of photons due to the squeezing effect. For a mode of constant frequency $\omega_1$, the variation of work vanishes because
\begin{equation}
\delta W_1 = \frac{d \omega_1}{d t} (\sinh^2 r(t) + \frac{1}{2}) = 0 .
\end{equation}
The variation of heat is however different from zero. It reads
\begin{equation}
\delta Q = \omega_1 \sinh2r(t) \frac{\partial r(t)}{\partial t} d t .
\end{equation}
It can also be checked that 
\begin{equation}
\sigma \equiv \frac{d S_{ent}}{d t} - \frac{1}{T(t)} \frac{\delta Q}{\delta t}  = 0 \; \; , \; \forall t .
\end{equation}
Therefore, the second principle of thermodynamics provides us with no arrow of time because the entropy production $\sigma$ identically vanishes at any time. In a non-reversible process, however, the constraint $\sigma > 0$ would give rise to the \emph{entanglement thermodynamical arrow of time} \cite{Partovi2008, Jennings2010}.

\section{Quantum entanglement in the multiverse}

\subsection{Creation of entangled pairs of universes}

First, we shall present a plausible scenario for the nucleation of a pair of entangled universes. The Wheeler-de Witt equation (\ref{WDW1}) for a de-Sitter universe with a massless scalar field reads 
\begin{equation}\label{WDW3}
\hbar^2 \ddot{\phi} + \frac{\hbar^2}{a}  \dot{\phi} -\frac{\hbar^2}{a^2} \phi'' + (\Lambda a^4 - a^2) \phi = 0 ,
\end{equation}
where, $\phi\equiv \phi(a,\varphi)$ is the wave function of the universe with, $\dot{\phi}\equiv\frac{\partial \phi}{\partial a}$ and $\phi'\equiv\frac{\partial \phi}{\partial \varphi}$, and $\Lambda$ is the cosmological constant. As it was already pointed out in Sec. 2, in the third quantization formalism the wave function $\phi$ is promoted to an operator $\hat{\phi}$ that, in the case now being considered, can be decomposed in normal modes as
\begin{equation}\label{decomposition}
\hat{\phi}(a,\varphi) = \int d k \, \left( e^{i k \varphi} A_k(a) \hat{b}_k^\dag + e^{-i k\varphi} A^*_k(a) \hat{b}_k \right) ,
\end{equation}
where, $\hat{b}_k\equiv \hat{b}_k(a_0)$ and $\hat{b}_k^\dag\equiv \hat{b}_k^\dag(a_0)$, are the constant operators defined in Eqs. (\ref{b01}-\ref{b02}), now with the mode-dependent frequency,
\begin{equation}\label{Frequency0}
\omega_k(a) = \frac{1}{\hbar}\sqrt{\Lambda a^4 - a^2 + \frac{\hbar^2 k^2}{a^2}} ,
\end{equation}
evaluated at $a_0$. The probability amplitudes $A_k(a)$ and $A_k^*(a)$ satisfy the equation of the damped harmonic oscillator,
\begin{equation}\label{equationModes}
\ddot{A}_k(a) + \frac{\dot{\mathcal{M}}}{\mathcal{M}} \dot{A}_k(a) + \omega_k^2 A_k(a) = 0 ,
\end{equation}
with, $\mathcal{M}\equiv \mathcal{M}(a) = a$, and $\omega_k\equiv\omega_k(a)$. Let us recall that the real values of the frequency (\ref{Frequency0}) define the oscillatory regime of the wave function of the universe in the Lorentzian region, and the complex values define the exponential regime of the Euclidean region. Let us first consider the zero mode of the wave function, i.e. $k=0$. Then, the wave function $\phi_\Lambda(a)$ quantum mechanically describes the nucleation of a de-Sitter universe from a de-Sitter instanton  \cite{Vilenkin1982, Hawking1983, Vilenkin1986, Kiefer2007} depicted in Sec. 2, with a transition hypersurface $\Sigma_0 \equiv \Sigma(a_0)$ located at $a_0 =  \frac{1}{\sqrt{\Lambda}}$ (see, Fig \ref{DS_instanton}).

\begin{figure}[htb]
\begin{minipage}[b]{0.48\linewidth}
\centering
\includegraphics[width=4cm,height=3cm]{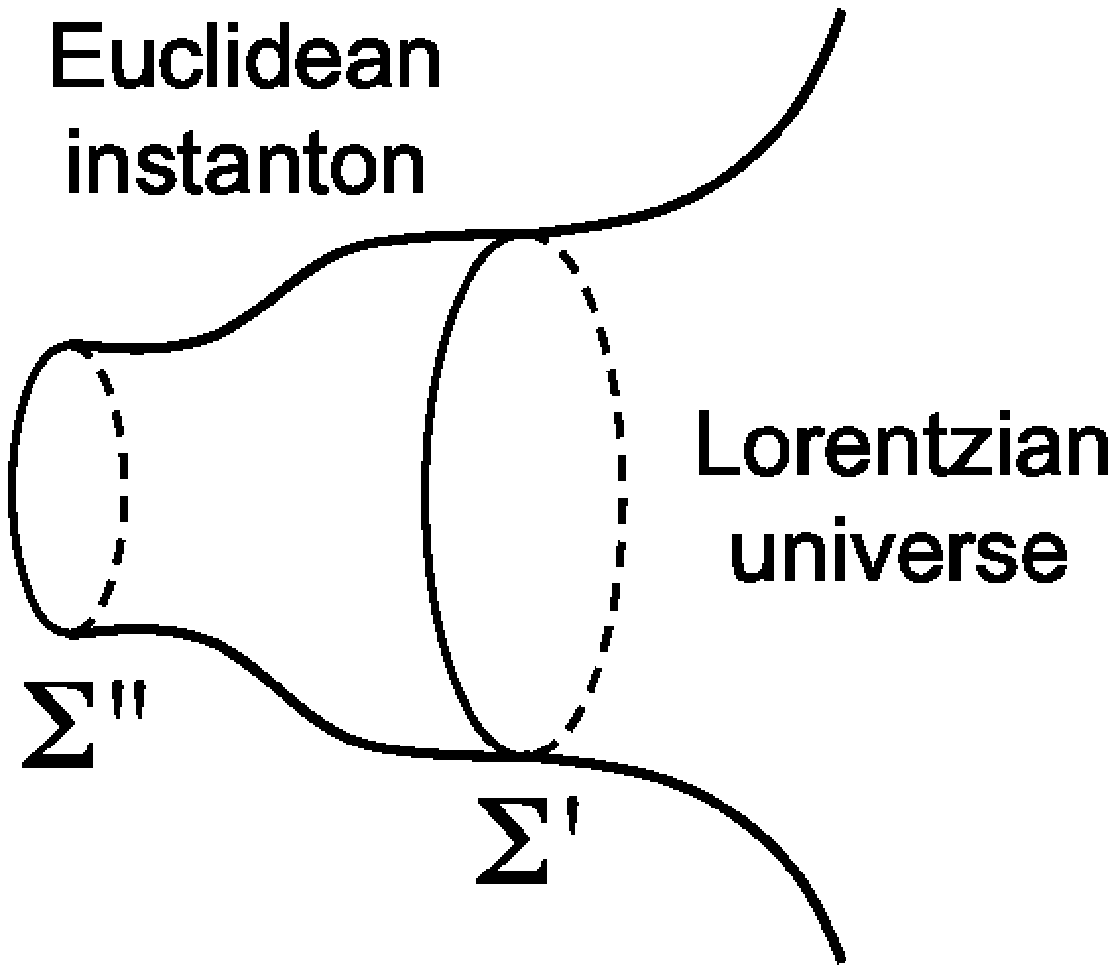}
\caption{Before reaching the collapse, the instanton finds the transition hypersurface $\Sigma''$.}
\label{EE_instanton}
\end{minipage}
\begin{minipage}[b]{0.52\linewidth}
\centering
\includegraphics[width=6cm,height=3cm]{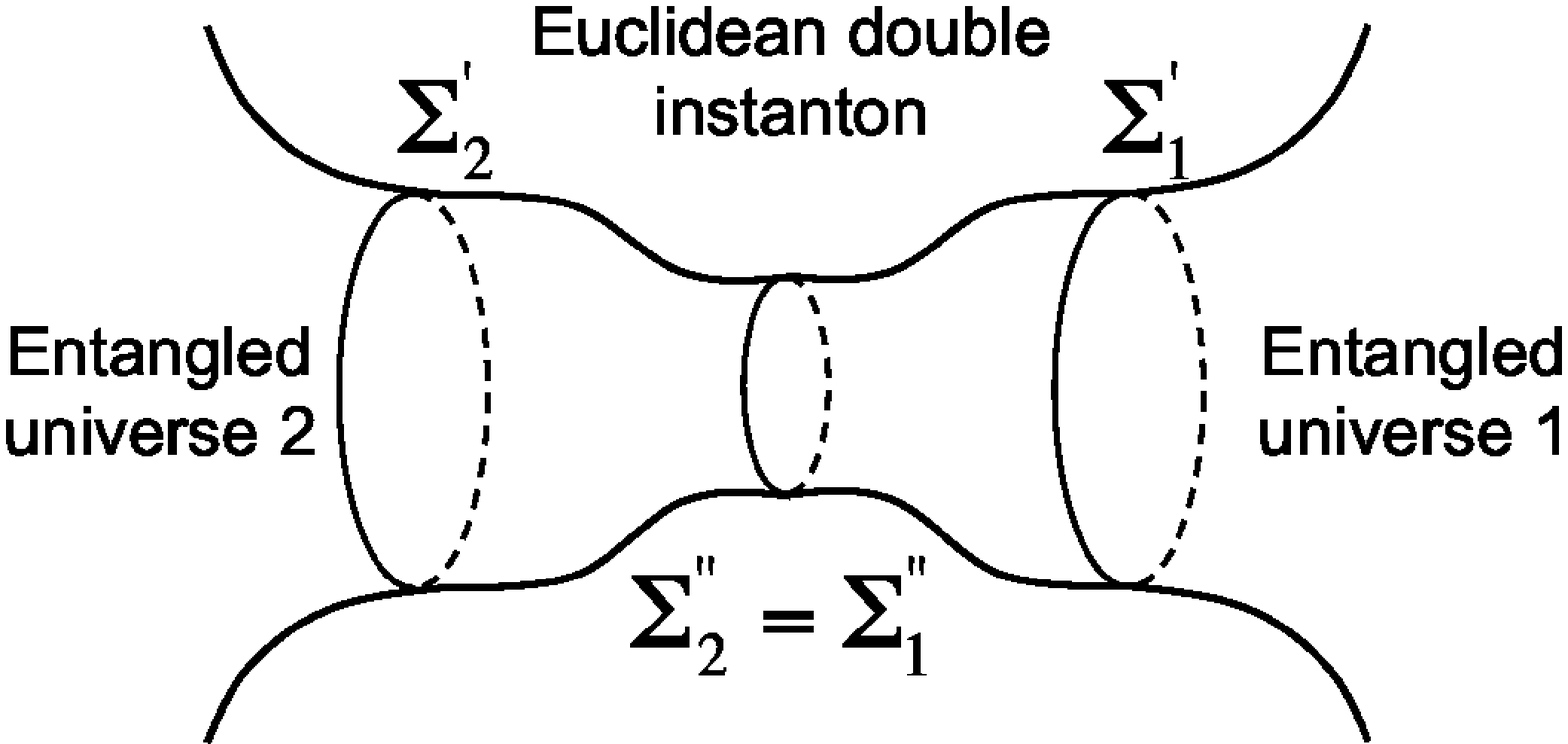} 
\caption{Creation of a pair of entangled universes from a pair of instantons.}
\label{EntangledInstanton}
\end{minipage}
\end{figure}

For values of $k$ different from zero, the quantum correction term given in Eq.  (\ref{Frequency0}) introduces a novelty. For the value, $k_{m} >  k > 0$, where $k_{m}^2\equiv \frac{4}{27 \hbar^2 \Lambda^2}$, there are two transition hypersurfaces from the Euclidean to the Lorentzian region, $\Sigma' \equiv \Sigma(a_+)$ and $\Sigma'' \equiv \Sigma(a_-)$, respectively, located at \cite{RP2011b}
\begin{eqnarray}\label{amas}
a_+ & \equiv & \frac{1}{\sqrt{3 \Lambda}} \sqrt{1 + 2 \cos\left(\frac{\theta_k}{3}\right)} , \\ \label{amenos}
a_- & \equiv & \frac{1}{\sqrt{3 \Lambda}} \sqrt{1 - 2 \cos\left(\frac{\theta_k+\pi}{3}\right) } , 
\end{eqnarray}
where, in units for which $\hbar=1$,
\begin{equation}\label{thetak}
\theta_k \equiv \arctan\frac{2 k \sqrt{k_{m}^2 - k^2 } }{k_{m}^2 - 2 k^2} .
\end{equation}
The picture is then rather different from the one depicted in Fig. \ref{DS_instanton}. First, at the transition hypersurface $\Sigma'$ the universe finds the Euclidean region (let us notice that for $k\rightarrow 0$, $a_+ \rightarrow a_0$ and $a_-\rightarrow 0$). However, before reaching the collapse, the Euclidean instanton finds a new  transition hypersurface $\Sigma''$  (see Fig \ref{EE_instanton}). Then, following a mechanism that parallels that proposed by Barvinsky and Kamenshchik in Refs. \cite{Barvinsky2006, Barvinsky2007a, Barvinsky2007b}, two instantons can be matched by identifying their hypersurfaces $\Sigma''$ (see Fig. \ref{EntangledInstanton}). The instantons can thus be created in pairs which would eventually give rise to an entangled pair of universes. Let us notice that this is a quantum effect having no classical analog because the quantum correction term in Eq. (\ref{Frequency0}) does not appear in the classical theory. 

The matching hypersurface $\Sigma'' \equiv \Sigma''(a_-)$, where $a_-\equiv a_-(\theta_k)$ is given by Eq. (\ref{amenos}) with Eq. (\ref{thetak}),  depends on the value $k$ of the mode.  Therefore, the matched instantons can only be joined for an equal value of the mode of their respective scalar fields. The  universes created from such a double instanton are then entangled, with a composite quantum state given by
\begin{equation}\label{EntInstanton}
\phi_{I,II} = \int dk \left( e^{ i k (\varphi_I +\varphi_{II})} A_{I,k} (a) A_{II,k}(a) \, \hat{b}_{I,k}^\dag \hat{b}_{II,k}^\dag + e^{ - i k (\varphi_I +\varphi_{II})} A_{I,k}^*(a) A_{II,k}^*(a) \, \hat{b}_{I,k} \; \hat{b}_{II,k} \right) ,
\end{equation}
where $\varphi_{I,II}$ are the values of the scalar field of each single universe, labelled by $I$ and $II$, respectively. The cross terms like $A_{I,k}  A^*_{II,k}$ cannot be present in the state of the pair of universes because the orthonormality relations between the modes \cite{RP2011b}. Then, the composite quantum state must necessarily be the entangled state represented by Eq. (\ref{EntInstanton}).

It is also worth mentioning that, in the model being considered, there is no Euclidean regime for values $k \geq k_m$ and, therefore, no universes are created from the space-time foam with such values of the mode. Then,  $k_m$ can  be considered the natural cut-off of the model. Let us also note that a similar behavior of the modes of the universe would be obtained for a non-massless scalar field provided that the potential of the scalar field, $V(\varphi)$, satisfies the boundary condition \cite{Kiefer2005, Kiefer2007}, $V(\varphi)\rightarrow 0$ for $a\rightarrow 0$.

\subsection{Entangled and squeezed states in the multiverse}

Entangled states, like those found in the preceding section or those appearing in the phantom multiverse \cite{PFGD2007, RP2011b}, can generally be posed in the quantum multiverse. Furthermore, the canonical representations of the harmonic oscillator that represent the quantum state of the multiverse, in the model described in Sec. 2, are related by squeezed transformations \cite{Kim2001}. Thus, squeezed states may generally be considered in the quantum multiverse.

As we saw in Sec. 3, entangled and squeezed states are usually dubbed 'non-classical' states because they are related to the violation of classical inequalities. Such violation is fundamentally associated to the complementary principle of quantum mechanics. In the multiverse, squeezed and entangled states may also violate the classical inequalities \cite{RP2011b}. However, the conceptual meaning of such violation can be quite different from that given in quantum optics. For instance, if the existence of entangled and squeezed states would imply a violation of Bell's inequalities, then, it could not be interpreted in terms of locality or non-locality because these concepts are only well-defined inside a universe, where space and time are meaningful. In the quantum multiverse, there is generally no common space-time among the universes and, therefore, the violation of Bell's inequalities would be rather related to the interdependence of the quantum states that represent different universes of the multiverse.

Like in quantum optics \cite{Vedral2006}, the violation of the classical inequalities in the multiverse depends on the representation which is chosen to describe the quantum states of the universes \cite{RP2011b}. Unlike quantum optics, we do not have an experimental device to measure other universes rather than our own universe\footnote{In some sense, we are the 'measuring device' of our universe.}. However, the extension of the complementary principle to the quantum description of the multiverse  entails two main consequences. On the one hand, if the wave function of the universe has to be described in terms of 'particles', it means that in some appropriate representation we can formally distinguish the universal states as individual entities, giving rise therefore to the multiverse scenario. On the other hand, if it has to be complementary described in terms of waves, then, interference between the quantum states of two or more universes can generally be considered as well.

\subsection{Thermodynamical properties of entangled universes}

Let us consider a multiverse made up of homogeneous and isotropic universes with a slow-varying scalar field $\varphi$, recalling that in the case for which $\dot{\varphi}=0$ and $V(\varphi_0)\equiv \Lambda$, the model effectively represents a multiverse formed by de-Sitter universes.

Let us  consider one type of universes and describe the quantum state of the multiverse in terms of the annihilation and creation operators $\hat{b}(a)$ and $\hat{b}^\dag(a)$ given in Eqs. (\ref{b1}-\ref{b2}). The vacuum state of the multiverse, $|\bar{0}\rangle$, is then defined as the eigenstate of the annihilation operator $\hat{b}(a)$ with eigenvalue zero, i.e. $\hat{b}(a) |\bar{0}\rangle \equiv 0$. On the other hand, observers inhabiting a large parent universe would quantum mechanically describe the state of their respective universes in the asymptotic representation given by Eqs. (\ref{properRepresentation1}-\ref{properRepresentation2}), with a ground state $|0\rangle$ defined by, $\hat{b}_\omega(a) |0\rangle \equiv 0$. 

We can consider therefore two representations: the one derived from a consistent formulation of the boundary condition of the whole multiverse, or \emph{invariant representation}, given by the operators $\hat{b}(a)$ and $\hat{b}^\dag(a)$, and the asymptotic representation given by the operators $\hat{b}_\omega(a)$ and $\hat{b}_\omega^\dag(a)$, which might be called the \emph{observer representation}. They both are related by the squeezing transformation
\begin{eqnarray}
\hat{b} &=& \mu_\omega \, \hat{b}_\omega + \nu_\omega \, \hat{b}^\dag_\omega , \\
\hat{b}^\dag &=& \mu_\omega^* \, \hat{b}^\dag_\omega + \nu_\omega^* \, \hat{b}_\omega ,
\end{eqnarray}
where, $\mu_\omega\equiv \mu_\omega(a, \varphi)$ and $\nu_\omega\equiv \nu_\omega(a, \varphi)$, are given by
\begin{eqnarray}\label{muomega}
\mu_\omega(a,\varphi) &=& \frac{1}{2 \sqrt{\mathcal{M}(a) \omega(a, \varphi)}} \left( \frac{1}{R} + R \mathcal{M}(a) \omega(a, \varphi) - i \dot{R} \right) , \\ \label{nuomega}
\nu_\omega(a,\varphi) &=& \frac{1}{2 \sqrt{\mathcal{M}(a) \omega(a, \varphi)}} \left( \frac{1}{R} - R \mathcal{M}(a) \omega(a, \varphi) - i \dot{R} \right) ,
\end{eqnarray}
with $|\mu_\omega |^2 - |\nu_\omega|^2 = 1$, and $R\equiv R(a,\varphi)$ is given, in the semiclassical regime, by Eq. (\ref{R}),
\begin{equation}\nonumber
R \approx \frac{e^{\pm \frac{1}{3 V(\varphi)}}}{\sqrt{\mathcal{M}(a) \omega(a,\varphi)}} ,
\end{equation}
where the positive sign corresponds to the choice of the no-boundary condition and the negative sign to the tunneling boundary condition. Let us further assume that the multiverse is in the invariant vacuum state $|\bar{0}\rangle$. The density matrix that represents the quantum state of the multiverse turns out to be then
\begin{equation}
\hat{\rho}(a, \varphi) \equiv |\bar{0}\rangle \langle \bar{0}| =  \hat{\mathcal{U}}^\dag_S |0_1 0_2\rangle \langle 0_1 0_2 | \hat{\mathcal{U}}_S ,
\end{equation}
where $|0_1 0_2\rangle \equiv |0_1\rangle |0_2\rangle$, with $|0_1\rangle$ and $|0_2\rangle$ being the ground states of a pair of entangled universes in their respective observer representations. Similarly to Eq. (\ref{twomodeS}), the squeezing operator $\hat{\mathcal{U}}_S$ is given by  \cite{RP2011b}
\begin{equation}
\hat{\mathcal{U}}_S(a, \varphi) = e^{r(a, \varphi)  \hat{b}_1 \hat{b}_2 - r(a, \varphi)  \hat{b}_1^\dag \hat{b}_2^\dag} ,
\end{equation}
where the squeezing parameter, $r(a, \varphi)$, reads
\begin{equation}
r(a, \varphi) \equiv {\rm arcsinh} |\nu_\omega(a,\varphi)| ,
\end{equation}
with $\nu_\omega(a,\varphi)$ being given by Eq. (\ref{nuomega}). We can then follow the procedure of Sec. 3.3 to compute the reduced density matrix, $\hat{\rho}_1$, that represents the quantum state of one single universe of the entangled pair. It is given then by the thermal state \cite{RP2011b}
\begin{equation}\label{ThS}
\hat{\rho}_1(a,\varphi) \equiv {\rm Tr}_2\hat{\rho} = \frac{1}{Z} \sum_{N=0}^\infty e^{-\frac{\omega(a,\varphi)}{T}(N + \frac{1}{2})} |N\rangle \langle N|,
\end{equation}
with, $|N\rangle \equiv |N\rangle_2$ and $Z^{-1} = 2 \sinh\frac{\omega}{2 T}$. The two universes of the entangled pair evolve, in the observer representation of each single universe, in thermal equilibrium with a temperature of entanglement given by
\begin{equation}\label{eq6327}
T \equiv T(a,\varphi) = \frac{\omega(a,\varphi)}{2 \ln\frac{1}{\Gamma(a,\varphi)}}  ,
\end{equation}
where, $\Gamma(a,\varphi) \equiv \tanh r(a,\varphi)$.  The total energy reads
\begin{equation}\label{Etotal}
E(a) = \omega(a) (\langle N \rangle + \frac{1}{2}) ,
\end{equation}
where, $\langle N \rangle \equiv |\nu_\omega|^2$. The variation of the quantum informational analogues to the work, $W$, and heat, $Q$, now read 
\begin{eqnarray} \label{dW}
\delta W &=& \delta \omega \, (\langle N \rangle + \frac{1}{2}) \approx \frac{\partial \omega(a,\varphi)}{\partial a} (\langle N \rangle + \frac{1}{2}) \, da , \\ \label{dQ}
\delta Q &=& \omega   \, \delta \langle N \rangle \approx \omega(a,\varphi) \frac{\partial \langle N \rangle}{\partial a} \, da ,
\end{eqnarray}
where in the last equalities it has been taken into account that for a slow-varying field, $\delta\omega \approx \dot{\omega} \, da$ and $\delta\langle N \rangle\approx \dot{\langle N \rangle} \, da$. From Eqs. (\ref{Etotal}-\ref{dQ}) it can be checked that the first principle of thermodynamics, $\delta E = \delta W + \delta Q$, is directly satisfied. The entropy of entanglement, Eq. (\ref{entropy00}), reads 
\begin{equation}\label{eq6329}
S_{ent}(a, \varphi) = |\mu_\omega(a,\varphi)|^2 \, \ln |\mu_\omega(a,\varphi)|^2 - |\nu_\omega(a,\varphi)|^2 \, \ln |\nu_\omega(a,\varphi)|^2 ,
\end{equation}
with, $|\mu_\omega(a,\varphi)| = \cosh r(a,\varphi)$ and $|\nu_\omega(a,\varphi)| = \sinh r(a,\varphi)$. Therefore, like in Sec. 3.3, the second principle of thermodynamics is also satisfied because the entropy production vanishes for any values of the scale factor and the scalar field, i.e. $\sigma \equiv \sigma(a,\varphi)=0$.

\subsubsection{Entropy of entanglement as an arrow of time for single universes}

Let us summarize the general picture described so far. The multiverse stays in a squeezed vacuum state which is the product state of the wave functions that correspond to the state of pairs of entangled $i$-universes (see Eq. (\ref{stateMultiverse2})), where the index $i$ labels all the species of universes considered in the multiverse. The multiverse stays therefore in a highly non-classical state. Furthermore, the quantum entropy of a pure state is zero and, therefore, there is no thermodynamical arrow of time in the multiverse. Let us recall that, in the third quantization formalism, the scale factor was just taken as a formal time-like variable given by the Lorentzian structure of the minisupermetric. However, the minisuperspace is not space-time and, therefore, the scale factor has no meaning of a physical (i.e. a measurable) time, a priori, in the multiverse. It might well be said that (physical) time and (physical) evolution are concepts that really make sense within a single universe.

\begin{figure}
\begin{center}
\includegraphics[width=7cm,height=5cm]{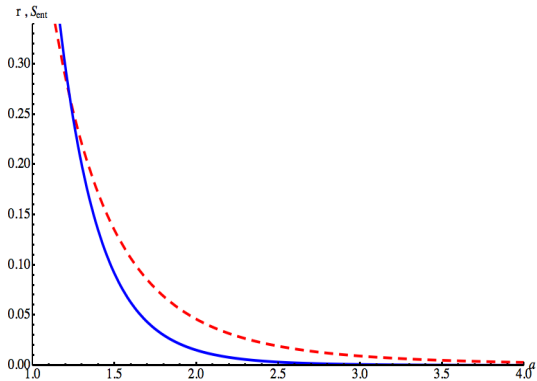}                

\caption{Parameter of squeezing, $r$ (dashed line),  and entropy of entanglement, $S_{ent}$ (continuous line), with respect to the value of the scale factor, $a$.}
\label{fig:entanglement}
\end{center}
\end{figure}

For an observer inside a universe, this is described by a thermal state which is indistinguishable from a classical mixture (see Eq. (\ref{thermal00}), and the comments thereafter), i.e. it is seen as a classical universe. The entropy of entanglement for a single universe is a monotonic function of the scale factor. However, the entropy production identically vanishes for any increasing or decreasing rate of the scale factor so that the customary formulation of the second principle of thermodynamics does not impose any arrow of time in the universe within the present approach. Although the universe can be seen as a classical mixture by an observer inside the universe, its quantum state has been obtained from a highly non-classical state. Thus, it would not be expected that the classical constraint $\sigma\geq 0$ would impose any arrow of time in the model. 

The second principle of entanglement thermodynamics \cite{Plenio1998} does provide us with an arrow of time for single universes. In the quantum multiverse, it can be reparaphrased as follows: by local operations and classical communications alone, the amount of entanglement between the universes cannot  increase. Let us recall that by \emph{local} operations we mean in the multiverse anything that happens within a single universe, i.e. everything we can observe. Therefore, the growth of cosmic structures, particle interactions and even the presence of life in the universe cannot increase the amount of entanglement between a pair of entangled universes provided that all these features are due to local interactions. They should  decrease the rate of entanglement in a non-reversible universe with dissipative processes, actually.

The amount of entanglement between the pair of universes only decreases for growing values of the scale factor (see Fig. \ref{fig:entanglement}). Thus, the second law of the entanglement thermodynamics implies that the universe has to expand once it is created in an entangled pair, as seen by an observer inside the universe. Furthermore, if the classical thermodynamics and the thermodynamics of entanglement were related, it could be followed that the negative change of entropy would be balanced by the creation of cosmic structures and other local processes that increase the local (classical) entropy. The decrease of the entropy of entanglement is larger for a small value of the scale factor. Then, the growth of local structures in the universe would be favored in the earliest phases of the universe, as it is expected.

\subsubsection{Energy of entanglement and the vacuum energy of the universe}

In the model being considered, $\sigma \equiv d S - \frac{\delta Q}{T}$ =0, and thus, the variation of the entropy of entanglement is related to the quantum informational heat, $Q$, by
\begin{equation}\label{Entropy01}
d S = \frac{\delta Q}{T}  .
\end{equation}
Eq. (\ref{Entropy01}) can be compared with the equation that is customary used to define the energy of entanglement \cite{Mukohyama1997, Mukohyama1998, Muller1995, Lee2007}, $dE_{ent} = T dS$. Then, in the case being considered, we can identify the energy of entanglement, $E_{ent}$, with the informational heat, $Q$, and  interpret  it as a vacuum energy for each single universe of an entangled pair. It is given by the integral of Eq. (\ref{dQ}), with 
\begin{equation}
\langle N \rangle \approx \frac{\, 9 \, e^{\pm\frac{2}{3 V(\varphi)}} }{16 V(\varphi)} \, \frac{1}{a^8}+ \sinh^2\frac{1}{3 V(\varphi)} ,
\end{equation}
where it has been used that, $\mathcal{M}(a)=a$ and $\omega = a\sqrt{a^2 V(\varphi)-1} \approx a^2\sqrt{V(\varphi)}$, in units for which $\hbar=1$. For a slow-varying field, $\varphi \approx \varphi_0$ and  $\delta \langle N \rangle \approx \frac{\partial \langle N \rangle}{\partial a} d a$, and therefore
\begin{equation}
\delta Q \approx \omega \frac{\partial \langle N \rangle}{\partial a} d a =- \frac{9 \, e^{\pm\frac{2}{3 V(\varphi_0)}}}{2 \sqrt{V(\varphi_0)}} \, a^{-7} d a ,
\end{equation}
whose integration yields
\begin{equation}\label{Eentanglement}
E_{ent} = Q(a,\varphi_0) =\frac{3}{4}  \frac{e^{\pm\frac{2}{3 V(\varphi_0)}}}{\sqrt{V(\varphi_0)}} \, a^{-6} .
\end{equation}
The energy of entanglement (\ref{Eentanglement}) provides us with a curve that might be compared with the evolution of the vacuum energy of the universe. From Eq. (\ref{Eentanglement}), it can be seen that the vacuum energy would follow a different curve depending on whether  the tunneling condition or the no-boundary condition is imposed on the state of a single universe. The boundary condition imposed on a single universe might therefore be discriminated from observational data, at least in principle. However, the model being considered is unrealistic for at least two reasons. First, after the inflationary stage the universe becomes hot \cite{Linde2007, Linde2008} and the slow-roll approximation is no longer valid. Secondly, if the energy of entanglement is to be considered as a vacuum energy, it should have been considered as a variable of the model from the beginning. More realistic matter fields and the backreaction should be taken into account to make a first serious attempt to observational fitting. However, the important thing that is worth noticing is that the vacuum energy of entanglement might thus be tested as well as the whole multiverse proposal. Furthermore, different boundary conditions would provide us with different curves for the energy of entanglement along the entire evolution of the universe. Therefore, the boundary conditions of the whole multiverse might be tested as well by direct observation, which is a completely novel feature in quantum cosmology.

\section{Conclusions: the physical multiverse}

In this chapter, we have presented a quantum mechanical description of a multiverse made up of large and disconnected regions of the space-time, called universes, with a high degree of symmetry. We have obtained, within the framework of a third quantization formalism, a wave function that quantum mechanically represents the state of the whole multiverse, and an appropriate boundary condition for the state of the multiverse has allowed us to interpret it as formed by entangled pairs of universes. 

If universes were entangled to each other, then, the violation of classical inequalities like the Bell's inequalities could no longer be associated to the concepts of locality or non-locality because there is not generally a common space-time among the universes of the quantum multiverse. It would rather be related to the independence or interdependence of the quantum states that represent different universes. Furthermore, the complementary principle of quantum mechanics, being applied to the space-time as a whole, enhances us to:  i) look for an appropriate boundary condition for which universes should be described as individual entities  forcing us to consider a multiverse; and, ii) take into account as well interference effects between the quantum states of two or more universes.

For a pair of entangled universes, the quantum thermodynamical properties of each single universe have been computed. In the scenario of a multiverse made up of entangled pairs of universes, the picture is the following: the multiverse may state in the pure state that corresponds to the product state of the ground states derived from the boundary condition imposed on the multiverse, for each type of single universes. Then, the entropy of the whole multiverse vanishes and there is thus no physical arrow of time in the multiverse. For single universes, however, it appears an arrow of time derived from the entropy of entanglement with their partner universes.

The entropy of entanglement decreases for an increasing value of the scale factor. The second principle of thermodynamics is however satisfied because the process is non-adiabatic, in the quantum informational sense, and the entropy production is zero. In fact, the entropy production is zero for any increasing or decreasing rate of the scale factor, imposing therefore no correlation between the cosmic arrow of time, which is given by expansion or contraction rate of a single universe, and the customary formulation of the second principle of thermodynamics. This is in contrast to what it happens inside a single universe, where there is a correlation between the cosmic arrow of time and the entropy of matter fields \cite{Hawking1985, Kiefer1995}. Let us recall that the entropy of entanglement is a quantum feature having no classical analogue and, thus, it is not expected that it imposes an arrow of time through the customary, i.e. classical, formulation of the second principle of thermodynamics.

The second principle of entanglement thermodynamics, which states \cite{Plenio1998} that the entropy of entanglement cannot be increased by any \emph{local} operation and any classical communication alone, does impose an arrow of time on single universes \cite{RP2012}. It should be noticed that by \emph{local} we mean in the multiverse anything that happens in a single universe. Therefore, everything that we observe, i.e. the creation of particles, the growth of cosmic structures, and even life, cannot make the inter-universal entanglement to grow provided that all these processes are internal to a single universe. In an actual and non-reversible universe they should induce a decreasing of the entropy of inter-universal entanglement, enhancing therefore the expansion of the universe that would induce a correlation between the growth of cosmic structures and the entanglement arrow of time.

In the model presented in this chapter, the energy of entanglement between a pair of entangled universes provides us with a vacuum energy for each single universes. The energy of entanglement of the universe is high in the early stage of the universes  becoming very small at later times. That behavior might be compatible with an initial inflationary universe, for which a high value of the vacuum energy is assumed, that would eventually evolve to a state with a very small value of the cosmological constant, like the current state of the universe. However, it is not expected that such a simple model of the universe would fit with actual observational tests. A more realistic model of the universes that form the multiverse, in which genuine matter fields were considered, is needed to make a serious attempt of observational fitting. However, the fact that inter-universal entanglement provides us with testable properties of our universe opens the door to future developments that would make falseable the multiverse proposal giving an observational support to the quantum multiverse. 

In conclusion, the question of whether the multiverse is a physical theory or just a mathematical construction, derived however from the general laws of physics, holds on whether the existence of other universes may affect the properties of the observable universe. Inter-universal entanglement is a novel feature that supplies us with new explanations for unexpected cosmic phenomena and it might allow us to test the whole multiverse proposal. It will be the future theoretical developments and the improved observational tests what will make us to decide whether to adopt or deny such a cosmological scenario.

\section{Acknowledgements}

In memory of Prof. Gonz\'alez D\'{\i}az, who always encouraged us to adopt a fearless attitude as well in science as in everyday life. The author wishes to thank I. Garay for the revision of the text and his valuable comments and P. V. Moniz for his help and support.

\bibliographystyle{vancouver}
\bibliography{bibliography}

\end{document}